\documentclass[fleqn,usenatbib]{mnras}

\usepackage{newtxtext,newtxmath}

\usepackage[T1]{fontenc}
\usepackage{ae,aecompl}
\usepackage{booktabs}
\usepackage{multirow}
\usepackage{siunitx}
\usepackage{rotating}
\usepackage{caption}

\usepackage{graphicx}	
\usepackage{float}
\usepackage{amsmath}	
\usepackage{float}
\usepackage{bm}
\usepackage{ulem}

\DeclareCaptionFormat{cont}{#1 (cont.)#2#3\par}

\title{Dynamics of dwarf galaxies in $f(R)$ gravity}

\author[I. de Martino, A. Diaferio, L. Ostorero]{
	Ivan de Martino$^{1}$\thanks{E-mail: ivan.demartino@usal.es}, Antonaldo Diaferio$^{2,3}$, Luisa Ostorero$^{2,3}$
	\\
	$^{1}$ Universidad de Salamanca,Facultad de Ciencias.,F\'isica Te\'orica, Salamanca, Plaza de la Merced s/n. 37008, Spain \\
	$^{2}$  Dipartimento di Fisica, Universit\`a di Torino,  Via P. Giuria 1, I-10125 Torino, Italy \\
	$^{3}$  Istituto Nazionale di Fisica Nucleare (INFN), Sezione di Torino, Via P. Giuria 1, I-10125 Torino, Italy
}

\date{Accepted XXX. Received YYY; in original form ZZZ}

\pubyear{2021}

\begin{document}
	\label{firstpage}
	\pagerange{\pageref{firstpage}--\pageref{lastpage}}
	\maketitle
	
	\begin{abstract}
		We use the kinematic data of the stars in eight dwarf spheroidal galaxies to assess whether $f(R)$ gravity can fit  the observed profiles of the line-of-sight velocity dispersion of these systems without resorting to dark matter. Our model assumes that each galaxy is spherically symmetric and has a constant velocity anisotropy parameter $\beta$ and constant mass-to-light ratio consistent with stellar population synthesis models. 
		We solve the spherical Jeans equation that includes the Yukawa-like gravitational potential appearing in the weak field limit of $f(R)$ gravity, and a Plummer density profile for the stellar distribution. The $f(R)$ velocity dispersion profiles depends on two parameters: the scale length $\xi^{-1}$, below which the Yukawa term is negligible, and the boost of the gravitational field $\delta>-1$. 
		$\delta$ and $\xi$ are not universal parameters,   
		but their variation within the same class of objects is expected to be limited. The $f(R)$ velocity dispersion profiles fit the data with a value $\xi^{-1}= 1.2^{+18.6}_{-0.9}$ Mpc  for the entire galaxy sample. On the contrary, the values of $\delta$ show a bimodal distribution that picks at  
		$\overline{\delta}=-0.986\pm0.002$ and $\overline{\delta}=-0.92\pm0.01$. { These two values disagree at $6\sigma$ and suggest a severe tension for $f(R)$ gravity. It remains to be seen whether an improved model of the dwarf galaxies or additional constraints provided by the proper motions of stars measured by future astrometric space missions can return consistent $\delta$'s for the entire sample and remove this tension.}
	\end{abstract}
	

	
	\section{Introduction}
	
	The $\Lambda$ Cold Dark Matter ($\Lambda$CDM) model provides a theoretical framework capable of explaining the formation and evolution of cosmic structures  \citep{Planck2020-V,Planck2020-VI,Planck2020-VII}. A few tensions remain, however, including the fine tuning problem of the cosmological constant \citep{Weinberg1987, Weinberg1989,  Capolupo2017} or the tension on the Hubble constant  \citep{DiValentino2021}. 
	
	On the scale of galaxies, the $\Lambda$CDM model also suffers from some tensions, for example (i) the cusp/core problem, the discrepancy between the predicted steep mass density profile in the  central regions of dark matter halos,
	and the core suggested by the observations of low surface brightness galaxies and dwarf galaxies; (ii) the missing satellite problem, the discrepancy between the predicted large number of dark matter sub-halos within the dark matter halo of a Milky Way-like galaxy and the observed small number of Milky Way satellites; and (iii) the too-big-to-fail problem, the predicted existence of bright massive satellites that remain unobserved (for comprehensive reviews see, e.g., \citealt{Boylan_Kolchin2011, Bullock2017,DelPopolo2017,deMartino2020,Salucci2021}).
	
	Baryonic feedback has been invoked as a possible solution to these challenges on the scale of galaxies; however, the details of the role and efficiency of baryonic feedback remain debated  \citep[e.g.,][]{deMartino2020}, 
	and it might thus not provide the full solution. Indeed, these challenges  
	may also point out to a breakdown of the standard law of gravity, General Relativity (GR) and its Newtonian weak-field limit  
	\citep{deMartino2020}.  
	The $f (R)$ theories are among the simplest and most studied extensions of GR. In these theories, the Einstein-Hilbert Lagrangian is replaced by a more generic function, $f(R)$, of the Ricci scalar, $R$. These theories give rise to fourth-order field equations where the additional degrees of freedom can be recast as an effective dark energy fluid. We can thus explain the accelerated expansion of the Universe without resorting to an exotic, and yet to be understood, dark component (see for instance \citealt{Starobinsky1980,Li2007, Nojiri2007, Starobinsky2007,WayneHU2007,Amendola2008, Cognola2008,Miranda2009, Nojiri2011, deMartino2015, Nojiri2017, Lazkoz_2018}, and references therein).
	
	A $f(R)$ gravity theory can also mimic the dark matter component required to describe the dynamics of cosmic structure in standard gravity \citep{Capozziello2012}. When we assume  
	a Taylor expandable $f(R)$ Lagrangian, $f(R)$ gravity gives rise to two effects in the weak-field limit of a spherically symmetric metric: (1) a boost of the intensity of the gravitational field quantified by the parameter $\delta$ (the smaller is $\delta$ in the range $-1<\delta<0$, the larger is the intensity of the field compared to  GR); and (2) a Yukawa-like correction to the Newtonian gravitational potential, which becomes relevant on length scales larger than $\xi^{-1}$  
	\citep{Capozziello2012,Banerjee2017}. For each astrophysical system described by a local curvature $R$, $\delta$ and $\xi$ are linked to the coefficients of the Taylor expansion of  $f(R)$ around a value $R_0$ corresponding to the curvature of the { local} background. For a Minkowski background $R_0=0$,  we have $\delta=1+f'_0$ and $\xi=(f'_0/6f''_0)^{1/2}$, where $f'_0$ and $f''_0$ are the first two non-null coefficients of the Taylor expansion. {We recover GR and the weak-field Newtonian limit when  $\delta=0$ and $\xi\to \infty$, so that $f(R) = R$. At the scales of stellar systems, thanks to the Chameleon mechanism \citep{Khoury2004},  the Yukawa deviations from the Newtonian potential leave unaltered the  consistency of the observations with the classical tests of GR \citep{Capozziello2008}. On the other hand, at galactic and extra-galactic scales, the  Chameleon mechanism is less effective due to the lower density of the environment and $\delta$ and $\xi$ can thus assume other values \citep{Capozziello2007}. Therefore,  $\delta$ and $\xi$ are not universal parameters and we might expect that they assume different values in different systems, although in objects within the same class, e.g. dwarf galaxies or stellar systems, the variation should be limited.}
	
	The effectiveness of this $f(R)$ model at mimicking the dark matter component of the standard model  was extensively explored on different scales, { where we can assume $R_0=0$}. On the galaxy and galaxy cluster scales, where the  dynamical role of the dark matter is dominant in the standard model, the estimated value of $\delta$ is close to $-1$, as expected. For example, \citet{Napolitano2012} investigated the stellar kinematics of elliptical galaxies:  they used long-slit data and planetary nebulae data out to seven times the effective radius of three elliptical galaxies to fit the velocity dispersion profiles and found $\delta$ in the range $[-0.88,-0.75]$.  Similarly, \citet{Capozziello2009} found  $\delta$ in the range $[-0.95,-0.85]$ by modelling  the mass profile of  12  X-ray galaxy clusters. \citet{DeMartino2014} found $\delta$ in the range $[-1,-0.43]$  by modeling, under the assumption of hydrostatic equilibrium, the gravitational potential well of about 600 X-ray galaxy  clusters  from the stacked  Sunyaev-Zel'dovich (SZ)  profiles from the Planck maps  of the temperature anisotropies \citep{Planck2014}.  \citet{DeMartino2016} derived $\delta=-0.48 \pm 0.22 $ from the SZ profile of the Coma cluster. These two latter results showed the capability of the $\delta$ boosting  of the gravitational field intensity to also explain the SZ emission without resorting to dark matter.  
	On the galaxy and  galaxy-cluster scales,  $\xi^{-1}\sim 1 $~Mpc, which is { consistent with the lack of evidence, on smaller scales, of the existence of a fifth force and of the effects of the accelerated expansion of the Universe. }
	
	The Yukawa correction to the Newtonian potential was also tested on the scales of stellar systems, where dark matter is not relevant in the standard model and, in $f(R)$, $\delta$ is expected to be close to $0$, { thanks to the effectiveness of the Chamaleon mechanism}. \citet{DeLaurentis2013} set $\delta=0$ and, by using the  first time derivative of the orbital period of binary pulsars, showed that the second derivative of the $f(R)$ Lagrangian computed on the background must be compatible with zero with a precision $\sim0.08$. Additionally, 
	\citet{DeMartino2018} and \citet{DeLaurentis2018} described  the effect of the Yukawa-term of the orbital motion of stars around a massive object. Their modelling led to a narrowing of the viable volume of the parameter space of the Yukawa potential on these scales; they found $\delta=-0.01_{-0.14}^{+0.61}$  and set a lower limit on  the scale length $\xi^{-1}> 6000$ AU \citep{deMartino2021}. 
	
	Here, we  extend the above analyses of $\delta$ and $\xi$ by investigating the internal dynamics of dwarf spheroidal galaxies (dSphs).  We adopt a Jeans analysis (Sect.~\ref{sec:jeans_model})  to model the observed velocity dispersion profiles of dSphs with a Monte Carlo Markov Chain (MCMC) approach (Sects.~\ref{sec:data} and \ref{sec:results}). We then compare our results with those obtained by \citet{Capozziello2009} with galaxy clusters, which are self-gravitating systems with mass-to-light ratios  comparable to those of dSphs, when estimated in standard gravity (Sect.~\ref{sec:discussions}). We conclude in {the same Section.} 
	
	\section{Jeans analysis in {\it f(R)} gravity}\label{sec:jeans_model}

	{ In the following, we} remind the basic equations of our dynamical model (Sect.~\ref{sec:basic}), and the stellar distribution we assume (Sect.~\ref{sec:stardensity}). { We then} summarize the derivation of the $f(R)$ gravitational potential entering the Jeans equation { (Sect.~\ref{sec:Yukawa})}.
	
	\subsection{Basic equations}\label{sec:basic}
	
	We can constrain the gravity theory with
	the stellar kinematics of a dSph by assuming that the galaxy is in dynamical equilibrium, and  that the dSph is supported by the star velocity dispersion  \citep{Mateo1998}, with a negligible galaxy rotation. If we assume that the system is spherically symmetric, and that the stars  trace the gravitational potential,  the Jeans equation  relates the  phase-space distribution of the stars to the gravity theory through the acceleration term  $d\Phi(r)/dr$ \citep{Lokas2003,Mamon2005,Binney2008,Mamon2010}  
	\begin{equation}\label{eq:Jeans}
		\frac{d[\rho_*(r)\sigma_r^2(r)]}{dr} = -\rho_*(r)\frac{d\Phi(r)}{dr}-2{  \beta}\frac{\rho_*(r)\sigma_r^2(r)}{r},
	\end{equation}
	where $\rho_*(r)$ is the mass density profile of the stellar component, $\sigma_r(r)$  the radial velocity dispersion, $\Phi(r)$ the gravitational potential, and ${  \beta}\equiv 1\sigma_t^2/2\sigma_r^2$  the velocity anisotropy parameter. In general, $\beta$ is a function of radius, but in the case $\beta={\mathrm{const}}$ then Eq.~\eqref{eq:Jeans} has the simple solution  \citep{Lokas2003} 
	\begin{equation}
		\rho_*(r)\sigma_r(r)=r^{-2{  \beta}}\int_{r}^{\infty} \frac{d\Phi(x)}{dx}\rho_*(x)x^{2{  \beta}} \,dx \, .
	\end{equation}
	In dSphs, we do not measure $\sigma_r(r)$, but  the velocity dispersion  projected along the line of sight \citep{Binney2008}
	\begin{equation}
		\sigma^2_{\mathrm{los}} (R_{\rm p}) = \frac{2}{\Sigma_*(R_{\rm p})}\int_{R_{\rm p}}^{\infty}\biggl(1-{  \beta}\frac{R_{\rm p}^2}{r^2}\biggr) \frac{\sigma^2_r(r)\rho_*(r)}{(r^2-R_{\rm p}^2)^{1/2}} r \,dr\,,
	\end{equation}
	where $R_{\rm p}$ is the radius projected onto the sky, and $\Sigma_*(R_{\rm p})$ is the stellar surface mass density{, that we provide in Sect.~\ref{sec:stardensity}.}
	
	\subsection{Stellar density profile }\label{sec:stardensity}
	
	Following \citet{Walker2009d}, we  assume  that the stellar  surface mass density of the dSphs is described by the Plummer model
	\begin{equation}\label{eq:surfacePlummer}
		\Sigma_*(R_{\rm p}) = { \frac{M_*}{L_V}} \frac{L_V}{\pi r_{1/2}^2}\left(1+\frac{R_{\rm p}^2}{r_{1/2}^2}\right)^{-2}
	\end{equation}
	where $L_V$ is the  luminosity in the $V$-band,  ${M_*}/{L_V}$ is a constant stellar mass-to-light ratio, and $r_{1/2}$ is the radius enclosing half of the total stellar luminosity. The Abel transform links the surface mass density in Eq.~\eqref{eq:surfacePlummer} to the three-dimensional mass density profile
	\begin{equation}
		\rho_*(r) =  { \frac{M_*}{L_V}}\frac{3 L_{V}}{4\pi r_{1/2}^3}\left(1+\frac{r^2}{r_{1/2}^2}\right)^{-\frac{5}{2}}\,.
		\label{Plummer}
	\end{equation}
	Table \ref{tab:1} lists the dSph luminosity in the $V$-band, $L_V$,  the stellar mass-to-light ratio, $M_*/L_V$, and the half-light radius, $r_{1/2}$,  adopted for each {of the} dSph{s considered} in our analysis.

	\subsection{Gravitational potential in {\it f(R)} gravity} \label{sec:Yukawa}
	
	Here, we briefly summarize the main steps leading to the  gravitational potential arising in the weak field limit of the $f(R)$ gravity theory.
	The action for $f(R)$ gravity is \citep{DeFelice2010}
	\begin{equation}
		S=\frac{1}{16\pi G_{\mathrm{N}}}\int d^4x \sqrt{-g}[f\left(R\right)+\mathcal{L}_m]\, ,
	\end{equation}
	where $\mathcal{L}_m$ is the Lagrangian for matter fields, $f(R)$ is an arbitrary function of the Ricci scalar $R$, $g$ is the determinant of the metric tensor $g^{\mu\nu}$, and $G_{\mathrm{N}}$ is the gravitational constant. By varying the action with respect to the metric tensor $g^{\mu\nu}$, we obtain the field equations 
	\begin{equation}
		f'\left(R\right)R_{\mu\nu}-\frac{f(R)g_{\mu\nu}}{2}-\nabla_\mu \nabla_\nu f'(R)+g_{\mu\nu}\square f'(R)=\chi T_{\mu\nu}\, ,
		\label{3a}
	\end{equation}
	where $\chi = 8\pi G_{\mathrm{N}}$, $f'(R)={df(R)}/{dR}$, and $T_{\mu\nu}$ is the energy momentum tensor for the matter fields. Finally, the trace of the field equations is 
	\begin{equation} 
		3\square f'(R)+f'(R)R-2f(R)=\chi T\, , 
		\label{3b}
	\end{equation}
	where $T=g^{\mu\nu}T_{\mu\nu}$. It is worth noting that GR is recovered by setting $f(R)=R$ into the field equations and into their trace.
	
	Following \citet{Capozziello2012} and \citet{Banerjee2017}, we may assume that  the $f(R)$ Lagrangian is a Taylor expandable function around a background value $R_0$ of the curvature 
	\begin{eqnarray}
		f(R)= \sum_n \frac{f^n(R_0)}{n!} (R-R_0)^n = f_0+f'_0 R+f''_0 R^2+....\, ,
		\label{3c}
	\end{eqnarray}
	where $f^n(R)$ is the $n$-th derivative with respect to the Ricci scalar, and $f_0=f(R_0)$ corresponds to the cosmological constant term. Assuming that the background space-time is  Minkowskian, i.e. $R_0=0$, the lowest order of the perturbation expansion of the field equations returns $f_0=0$, while we can rewrite  the first derivative of the $f(R)$ Lagrangian as $f'_0=1+\delta$ where $\delta$ encodes the deviation from GR \citep{Capozziello2012}. The condition $\delta>-1$ must hold, otherwise gravity would be repulsive. 
	
	With the expansion of Eq.~\eqref{3c}, at the post-Newtonian order, { by assuming a spherically symmetric metric and thus ignoring the mixed terms of the metric}, one obtains
	\begin{align}
		&\left(\nabla^2-\xi^2\right)R^{(2)}=-\frac{8\pi G_N\xi^2}{1+\delta}\rho\,,\\
		&\nabla^2\left(\frac{f'_0}{4}g_{00}^{(2)}+\frac{f'_0}{4}g_{ii}^{(2)}+2f''_0 R^{(2)}\right)=-8\pi\rho G_{\mathrm{N}}\,\\
		& \nabla^2\left(f'_0g_{ii}^{(2)}+5f'_0g_{00}^{(2)}\right)=-64\pi\rho G_{\mathrm{N}}
	\end{align}
	where $R^{(2)}$, $g_{ii}^{(2)}$, and $g_{00}^{(2)}$ are the Ricci scalar and the components of the metric tensors up to order $\mathcal{O}(2)$, respectively, and  $\xi=\sqrt{{f'_0}/{6f''_0}}$.  $\xi$ defines the mass of the scalar field $R^{(2)}$, and the condition $f''_0>0$ must hold to avoid tachyonic instability \citep{DeFelice2010}.   The solution of  the above system of equations leads to the modified Poisson equation
	\begin{equation}\label{eq:poisson}
		\nabla^2\Phi(\textbf{r})=\frac{4\pi G_{\mathrm{N}}}{1+\delta}\rho(\textbf{r})-\frac{1}{6\xi^2}\nabla^2R^{(2)}\, ,
	\end{equation}
	whose solution  for a spherically symmetric system  is
	\begin{align}
		\Phi\left(\textbf{r}\right)=&-\frac{G_{\mathrm{N}}}{1+{  \delta}}\int\frac{\rho\left(\textbf{r}'\right)}{\mid\textbf{r}-\textbf{r}'\mid}d^3r'-\nonumber\\
		&\frac{G_{\mathrm{N}}}{3 (1+{  \delta})}\int\frac{\rho\left(\textbf{r}'\right)}{\mid\textbf{r}-\textbf{r}'\mid}e^{-{  \xi}\mid\textbf{r}-\textbf{r}'\mid}d^3r'\,.
		\label{3e}
	\end{align}
	The acceleration experienced by a test particle is thus
	\begin{align}
		-\frac{d\Phi}{dr} =& -I_0(r)I_1(r) - \frac{I_0(r)}{3{  \xi} }\left(1+{  \xi} r\right)e^{-{  \xi} r}I_2(r)\nonumber \\
		&-\frac{I_0(r)}{{ 3\xi}}\left[\sinh\left({  \xi} r\right)-{  \xi} r \cosh\left({  \xi} r\right)\right]I_3(r)
		\label{fR}
	\end{align}
	where
	\begin{align}
		& I_0(r) =\frac{4\pi G_{\mathrm{N}}}{(1+{  \delta})r^2}\,,\\
		& I_1(r) = \int_{0}^{r}r'^2\rho(r')dr'\,,\\
		& I_2(r) = \int_{0}^{r}r'\rho(r')\sinh\left({  \xi} r'\right)dr'\,,\\
		& I_3(r) = \int_{r}^{\mathcal{R}}r'\rho(r')e^{-{  \xi} r'}dr'\,.\label{eq:I3}
	\end{align}
	
	In our case, in Eq.~(\ref{eq:I3}), $\mathcal{R}$ is the size of the galaxy, that is defined as the radius where the baryonic mass density profile is 0.01 times the central density. Equation \eqref{3e} shows that, when $\delta<0$, the intensity of the gravitational field increases with respect to the Newtonian field; in addition, $\xi^{-1}$ represents the scale length beyond which the non-Newtonian behaviour starts dominating the dynamics of the self-gravitating system. We remind that, throughout our analysis, we  assume that no dark matter is present. Therefore, the mass density $\rho(r)$ appearing in the previous equations coincides with the stellar mass density $\rho_*(r)$ given in Eq. \eqref{Plummer}.
	
	\section{Data and data analysis}\label{sec:data}

	In our analysis, for each dSph we derive the values of the parameters $\delta$ and $\xi$ of the $f(R)$ gravity model, the velocity anisotropy parameter $\beta$, and the mass-to-light ratio $M_*/L_V$ by modelling the measured line-of-sight velocity dispersion profiles with the expected profiles obtained by solving the Jeans equation illustrated in Sect.~ \ref{sec:basic}. 
	In Sect.~\ref{sec:veldata} we describe the data set we adopt, and in Sect.~\ref{sec:M/L} we derive the expected mass-to-light ratio allowed by stellar population synthesis models. In Sect.~\ref{sec:mcmc} we illustrate our MCMC analysis.
	
	\subsection{Projected velocity dispersion profiles}\label{sec:veldata}
	
	We use the kinematic data sets of eight dSphs, namely Carina, Fornax, Sculptor, 
	Sextans, Draco, Leo I, Leo II, and Ursa Minor.
	The kinematic data sets of Carina, Fornax, Sculptor,
	and Sextans were obtained with the Michigan/MIKE Fiber Spectrograph (MMFS) at Magellan \citep{Walker2007, Walker2009a, Walker2009b, Walker2009c, Walker2009d}, while 
	the kinematic data sets of Draco, Leo I, Leo II, and Ursa Minor were obtained 
	with the Hectochelle fiber spectrograph at the MMT \citep{Mateo2008}.  For each dSph, Table \ref{tab:1} lists the dSph quantities relevant for our analysis, including the distance $D_\odot$ of the galaxy from the observer, the distance $D_p$  of the pericentre of the dSph orbit around the Milky Way from the Milky Way, and the value of the velocity anisotropy parameter $\beta_{\mathrm{NFW}}$  when  a Navarro-Frank-White density profile for the dark matter is adopted in standard gravity \citep{Navarro1996, Walker2009d}. Below, we compare $\beta_{\mathrm{NFW}}$ with the $\beta$ found in $f(R)$ gravity.
	
	Measuring the velocity dispersion profile requires the identification of the stars that are members of the dSph. \citet{Walker2009b} estimated a membership probability for each observed star by using an iterative expectation maximization technique on the position of the star, its magnesium index and its line-of-sight velocity. \citet{Walker2007} and \citet{Walker2009a, Walker2009c} derived the velocity dispersion profiles by including all the stars with membership probability larger than 95\%. The member stars are binned in radial circular annuli containing the same number of stars. The bulk transverse motion of the dSph is subtracted.
	
	dSphs move in the gravitational potential well of the Milky Way, and tidal effects may affect the kinematics of the outer stars. For instance, there is evidence of tidal stripping in Carina \citep{Munoz2008} and Leo I \citep{Sohn2007, Mateo2008}. On the contrary, tidal effects appear negligible in Fornax \citep{Piatek2007, Walker2009c}. 
	If we consider the dSph and the Milky Way as point  masses of mass $m$ and $M$, respectively, separated by  distance $D_p$ at the pericentre, in Newtonian gravity the tidal radius is 
	\begin{equation}\label{eq:rtidal}
		r_{\mathrm{tidal}}\approx D_p \biggl(\frac{m}{M}\biggr)^{1/3}\,,
	\end{equation}
	\citep{vonHoerner1957,King1962}. In the Yukawa potential, the tidal radius can be recast in the same form because the scale length $\xi^{-1} \gg D_p$. { In this case, $m$ and $M$ only include the baryonic mass, unlike Newtonian gravity, where $m$ and $M$ include both the baryonic mass and dark matter.   } 
	
	With the pericentre distance in the range $D_p \in (30\div60)$ kpc, depending on the dSph \citep{Fritz2018}, and a Milky Way mass $M=1.08_{-0.14}^{+0.20} \times 10^{12} M_\odot$, as inferred from the second data release of the {\it Gaia} satellite \citep{Cautun2020}, { in Newtonian gravity} we find the tidal radii $r_{\mathrm{tidal}}$ listed in Table \ref{tab:1}.  These radii are at least 30\% larger than the radii corresponding to the last data point of the velocity dispersion profile. The only exception is Draco: for this dSph the tidal radius almost coincides with the radius of the last data point. 
	
\begin{figure*}
		\includegraphics[width=2\columnwidth,keepaspectratio]{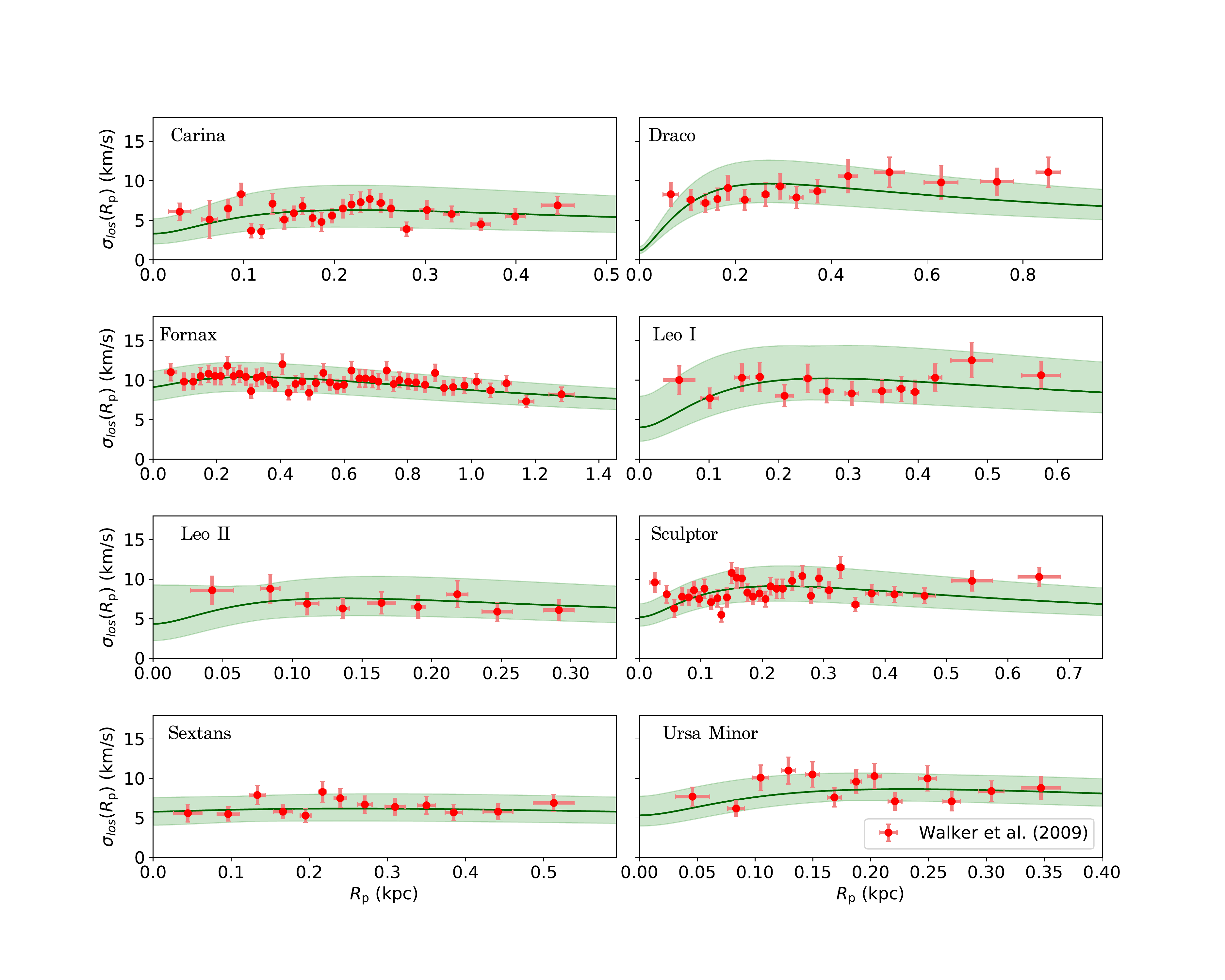}
		\caption{Radial profiles of the line-of-sight velocity dispersions of the eight dSphs { listed in Table \ref{tab:1}}. The red circles with error bars show the measured $\sigma_{\mathrm{los,\, obs}}(r_i)$ from \citet{Walker2009c}. The green solid lines show the model profiles $\sigma_{\mathrm{los,\, th}}(\bm{\theta},r_i)$ in $f(R)$ gravity adopting the best-fit parameters $\bm{\theta} = \{\delta, \xi, \beta, M_*/L\}$ listed in Table \ref{tab:2}; the green shaded areas show the  corresponding 1$\sigma$ spreads of the model profiles. }
		\label{fig:2}
	\end{figure*}

	{ We can adopt the same tidal radii in $f(R)$ gravity: indeed, the ratio of Newtonian masses including dark matter in Eq.~(\ref{eq:rtidal}) coincides with the ratio of the baryonic masses in $f(R)$ gravity, because the enhancement of the intensity of the gravitational field due to the Newtonian masses $m$ and $M$ including dark matter is mimicked by the masses $m/(1+\delta)$ and $M/(1+\delta)$ in $f(R)$ gravity, where now $m$ and $M$ are the baryonic masses alone. Therefore, both in Newtonian gravity and in $f(R)$ gravity,} the tidal disruption is negligible in the innermost part of the galaxies, $r< 1 $ kpc, where { the  measures are available.} Nevertheless, to be conservative, we  follow \citet{Walker2009b} and discard the outermost two data points of the velocity dispersion profile of each dSph. The measured velocity dispersion profiles are shown as solid red dots with error bars in Fig.~\ref{fig:2}.
	
	\begin{table*}
		\begin{center}
			\resizebox{14cm}{!}{
				\setlength{\tabcolsep}{4pt}
				\begin{tabular}{lcccccc||cc||ccc}
					\hline
					\hline
					Galaxy & $D_{\odot}$ &  $D_{p}$ &$\log(L_{\rm V})$  & $r_{1/2}$ & $M(<r_{\mathrm{last}})$  &Ref. & $r_{\mathrm{tidal}}$ & $\frac{r_{\mathrm{last}}}{r_{\mathrm{tidal}}}$ & $\langle B-V \rangle$ & $\langle V-I \rangle$ & $M_*/L_V$     \\
					& (kpc)                & (kpc)                &($L_{\odot}$)      & (pc)     & ($10^7 M_\odot$)  & & (kpc) & & \\
					(1)         & (2)      & (3)  & (4) & (5)  & (6) &  (7) & (8)   & (9)   & (10) & (11)  & (12) \\
					\hline
					\textbf{Carina} & 105$\pm$6  & $60^{+21}_{-16}$ & 5.57$\pm$0.20 & 273$\pm$45 & $3.7^{+2.1}_{-1.8}$  &[1, 2, 3, 15] & $\sim 2.0$ & $\sim 2.0$& - & $1.1\pm0.2$ & $3.4\pm2.9$ \\[0.1cm]
					\textbf{Draco} & 76$\pm$5  & $28^{+12}_{-7}$  &5.45$\pm$0.08 & 244$\pm$9 & $26.4^{+18.6}_{-17.4}$ &  [3, 4, 5, 6, 15] & $\sim 1.7$ & $\sim  1.0$ & - & $1.4\pm0.1$ & $11.1\pm4.7$ \\[0.1cm]
					\textbf{Fornax} & 147$\pm$12  & $69^{+26}_{-18}$  &7.31$\pm$0.12 & 792$\pm$58&  $12.8^{+2.2}_{-5.6}$ & [1, 2, 3, 15]& $\sim 2.8$ & $\sim 1.7$ & - & $1.3\pm0.2$ & $7.1\pm6.0$ \\[0.1cm]
					\textbf{Leo I} & $254^{+19}_{-16}$  & $45^{+80}_{-34}$ & 6.74$\pm$0.12 & 298$\pm$29& $8.9^{+4.3}_{-5.2}$&  [2, 3, 7, 8, 15] & $\sim 2.0$ & $\sim 2.1$ &  - & $1.4\pm0.1$ & $8.8\pm5.6$\\[0.1cm]
					\textbf{Leo II} & 233$\pm$15  & $45^{+121}_{-30}$   & 5.87$\pm$0.12 & 219$\pm$52 & $1.7^{+1.9}_{-1.2}$  &[2, 3, 9, 10, 15] & $\sim 1.1$ & $\sim 2.6$ & - & $0.6\pm0.2$ & $0.4\pm0.4$\\[0.1cm]
					\textbf{Sculptor} & 86$\pm$6  & $50^{+15}_{-10}$  &6.36$\pm$0.20 & 311$\pm$46 & $10.0^{+3.2}_{-5.0}$ &[2, 3, 11, 15] & $\sim 2.3$ & $\sim 2.1$&  - &$1.1\pm0.1$ &$3.6\pm2.0$ \\[0.1cm]
					\textbf{Sextans} & 86$\pm$4  & $71^{+11}_{-12}$  &5.64$\pm$0.20 & 748$\pm$66 & $2.0^{+1.0}_{-0.7}$ & [2, 3, 12, 15]& $\sim 1.9$&  $\sim 1.9$ & - & $1.4\pm0.1$ &  $8.5\pm3.3$\\[0.1cm]
					\textbf{Ursa Minor} & 76$\pm$4  & $29^{+8}_{-6}$  & 5.45$\pm$0.20 & 398$\pm$44 & $4.4^{+2.9}_{-2.0}$ &[2, 3, 13, 14, 15]& $\sim 1.0 $ & $\sim 1.3$ & $0.5\pm0.3$ &  - & $1.2\pm1.3$\\[0.1cm]
					\hline
				\end{tabular}
			}
		\end{center}
		\caption{Observational properties of the eight dSphs {analysed in this work}. Columns (2) and (3): distance of the dSph  from the observer and distance of the pericentre of the dSph orbit around the Milky Way from the Milky Way { center of mass}; Column (4): total $V$-band luminosity; Column (5): half-light radius; Column (6): total mass,  estimated by \citet{Walker2009c} assuming Newtonian gravity, within the
			outermost data point of the velocity dispersion profile; 
			Column (7):  sources of the observational data: [1] \citet{Pietrzynski2009}, [2] \citet{Irwin1995}, [3] \citet{Walker2009c}, [4] \citet{Bonanos2004}, [5] \citet{Martin2008}, [6] \citet{Walker2007}, [7] \citet{Bellazzini2004}, [8] \citet{Mateo2008}, [9] \citet{Bellazzini2005}, [10] \citet{Koch2007}, [11] \citet{Pietrzynski2008}, [12] \citet{Lee2009}, [13] \citet{Carrera2002}, [14] \citet{Walker2009b}, [15] \citet{Fritz2018}; Columns (8), and (9): tidal radius in Newtonian gravity [Eq. \eqref{eq:rtidal}] adopting the total Milky Way mass  $M=1.08_{-0.14}^{+0.20} \times 10^{12} M_\odot$ \citep{Cautun2020} and radius of the last data point of the velocity dispersion profile;  
			Column (10), and (11): average 
			values of the $B-V$ and $V-I$ indexes (see Sect.~\ref{sec:M/L}); Column (12): stellar mass-to-light ratios in the $V$-band (see Sect.~\ref{sec:M/L}).
		}\label{tab:1}
	\end{table*}

	\subsection{Expected mass-to-light ratios}\label{sec:M/L}
	
	In Newtonian gravity, the mass-to-light ratio of a dSph can be adjusted to any value depending on the mass associated to the dark matter halo embedding the stellar component. In our $f(R)$ model, where dark matter is absent, the mass-to-light ratio must be consistent with the value expected for the stellar population of the dSph. Nevertheless, in our MCMC analysis that we illustrate below, we still allow the mass-to-light ratio to vary as a free parameter within a limited range. Here, we derive the expected ratio around which we allow the mass-to-light ratio to vary in our MCMC analysis. 
	
	According to stellar population synthesis models \citep{Bell2001},
	\citet{Portinari2003, Portinari2004}  derived relations between the mass-to-light ratio and the colour of a stellar population. 
	We consider the following relations (Appendix B in \citealt{Portinari2004})
	\begin{align}\label{eq:b-v}
		& ( B - V ):\quad \log\biggl(\frac{M_*}{L_X}\biggr) = s_X [( B - V ) - 0.6] + q_X\,,\\
		& \label{eq:i-v} ( V - I ):\quad \log\biggl(\frac{M_*}{L_X}\biggr) = s_X [( V - I ) - 0.9] + q_X\,.
	\end{align}
	Here, the subscript $X$ indicates the magnitude band ($V$ in our case), and the coefficient $s_X$ is equal to 1.29 and 1.66 for  the $( B - V )$ and $( V - I )$ relations, respectively. The parameter $q_X$ is estimated by assuming a Kennicutt Initial Mass Function \citep{Kennicutt1983} in the range $[0.1-100]M_\odot$, and it is equal to 0.028 and -0.002 for the $( B - V )$ and $( V - I )$ relations, respectively. 
	
	We estimate the expected stellar mass-to-light ratio $M_*/L_V$ from the measured colour indexes of each dSph. 
	These colour indexes are the average of the colour indexes measured in different regions of the same galaxy. We adopt the spread of these measures as their uncertainties. 
	The colour indexes of Carina, Fornax, Sculptor, and Sextans are from  the Magellan/MMFS Survey \citep{Walker2009a}, the colour indexes of Leo I, Leo II, Draco, and Ursa Major are from \citet{Mateo2008,Lepine2011,Kinemuchi2008,Olszewski1985}, respectively. For each galaxy, we estimate the stellar mass-to-light ratio $M_*/L_V$ and its error by randomly sampling $N=1000$ colour indexes from a Gaussian distribution whose mean and width are the average index and its corresponding uncertainty. Equations \eqref{eq:b-v} or \eqref{eq:i-v} thus provide $N$ values of $M_*/L_V$. Table \ref{tab:1} lists the average of these  $M_*/L_V$ values and their spread.

	\subsection{ Monte Carlo Markov Chains}\label{sec:mcmc}
	
	Modeling the observed velocity dispersion profile $\sigma_{\mathrm{los,\, obs}}(r)$ measured by \citet{Walker2009d} with the profile $\sigma_{\mathrm{los,\, th}}(r)$  expected in $f(R)$ gravity requires the estimation of the { set of} four free parameters $\bm{\theta} =$ \{$\delta$, $\xi$, $\beta$, $M_*/L_V$\}. We explore this four-dimensional parameter space with the MCMC algorithm \texttt{emcee} \citep{emcee}.
	
	We assign a uniform prior distribution to the first three parameters in the ranges $\delta\in (-1;10]$, $\xi \in [0.01, 10^3]$ Mpc$^{-1}$, and $\beta\in [-100, 1)$. For each dSph, for the mass-to-light ratio $M_*/L_V$, we set a Gaussian prior  with mean and dispersion derived in Sect.~\ref{sec:M/L} and listed in Column (13) of Table \ref{tab:1}.
	We adopt the posterior probability  distribution 
	\begin{align}
		-2\log P(\bm{\theta}|\textrm{ data}) \propto& \sum_i\biggl[\frac{\sigma_{\mathrm{los,\, th}}(\bm{\theta},\, r_i)-\sigma_{\mathrm{los,\, obs}}(r_i)}{\Delta\sigma_{\mathrm{los,\, obs}}(r_i)}\biggr]^2\,,
		\label{eq:likelihood}
	\end{align}
	where $\Delta\sigma_{\mathrm{los,\, obs}}(r_i)$  are the observational uncertainties. We consider that our chains have converged when (1) all the chains are longer than 100 times  the  autocorrelation time, and (2) the autocorrelation time varies by less than 1\% (for more details we refer to Sect.~3 in \citealt{deMartino2022}).
	
	\section{Results} \label{sec:results}
	
	In our MCMC analysis, we employ 12 chains with random starting
	points selected from the prior distributions  described in Sect.~\ref{sec:mcmc}.
	Figure \ref{fig:1} shows the post burn-in posterior distributions of the parameters for each dSph derived with all these 12 chains. 
	{ We adopt the median of each posterior distribution as the best estimate of each of our free parameter, and the range within the 15.9 and 84.1 percentiles of each distribution as the 68\% confidence interval.}
	These values are listed in Table \ref{tab:2}.
	
	In Fig.~\ref{fig:2}, the green solid lines show the velocity dispersion profiles obtained with the parameters listed in Table \ref{tab:2}. The green shaded areas show the 1$\sigma$ uncertainties of the profiles. These uncertainties are the spread of 1,000 profiles derived by randomly sampling 1,000 times the input parameters within their posterior distribution obtained from our MCMC analysis.
	The model appears to properly describe  the observed profiles measured by \citet{Walker2009d} and shown by the red circles with error bars.
	
	In the standard model, the observed velocity dispersion profiles are modeled by assuming the presence of dark matter. In $f(R)$ gravity, the role played by the dark matter is played by the two parameters $\xi$ and $\delta$ listed in Table \ref{tab:2}. This table also lists the first two coefficients $f'_0$ and $f''_0$ of the Taylor expansion of the function $f(R)$\footnote{The uncertainties of $f'_0$ and $f''_0$  are the spread of these parameters derived by randomly sampling 1,000 times $\xi$ and $\delta$ within their posterior distributions obtained from our MCMC analysis. } (see Sect.~\ref{sec:Yukawa}). 
	For all the dSphs, the parameter $\xi$ appearing in the exponential term of the gravitational potential [Eq.~\eqref{3e}] is  $\xi \sim 1 $ Mpc$^{-1}$, and this Yukawa term is thus negligible on the scale of a few kiloparsec, corresponding to the size of the dSphs. It follows that modelling the kinematics of the stars in $f(R)$ gravity requires large departures of $\delta$ from the value $\delta=0$ of standard gravity. Indeed, for all the dSphs, the best value is close to the lower limit $\delta=-1$, and the value $\delta=0$ is excluded at $\sim 30\sigma$   for Carina, Draco, Leo I, and Sextans, and at $\sim 250\sigma$  for Fornax and Ursa Minor. 
	
	The absence of dark matter requires the mass-to-light ratio $M_*/L_V$ to be within the values expected from stellar population synthesis models. Our Gaussian priors set around these values guarantee that this conditions is verified, as shown in Fig.~\ref{fig:1}.
	The velocity field of the stars inferred in $f(R)$ gravity is similar to the field inferred in standard gravity: the velocity anisotropy parameter $\beta$ we derive is within $2.5\sigma$ at most from the values $\beta_{\mathrm {NFW}}$ derived by \citet{Walker2009d} assuming a dark matter halo with a NFW density profile in standard gravity. 
	
\begin{figure*}
		\includegraphics[width=2\columnwidth,keepaspectratio]{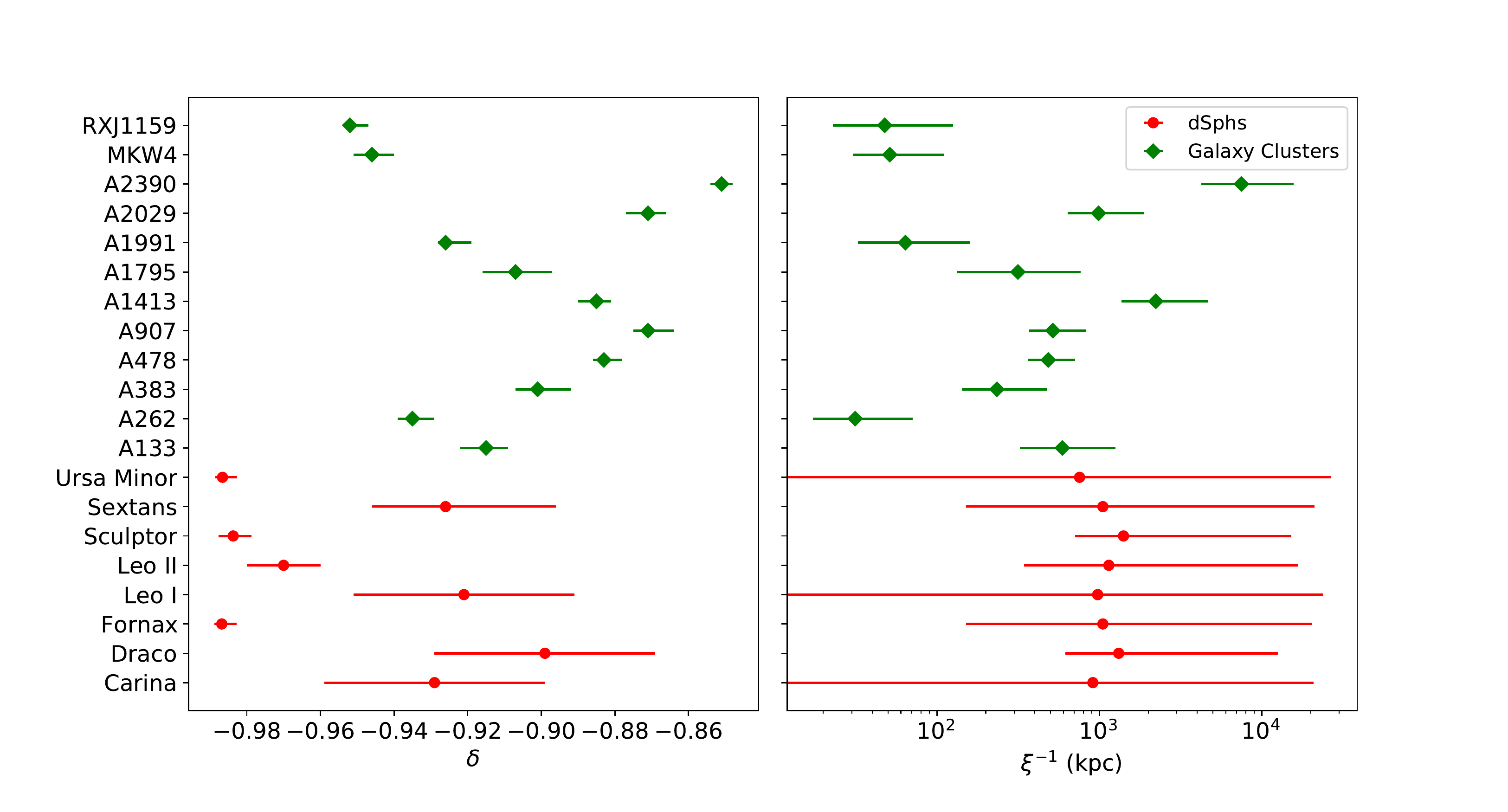}
		\caption{MCMC estimates of the parameters $\delta$ ({left panel}) and $\xi$ ({right panel}) and their 1$\sigma$ confidence intervals for the dSphs   { analysed in this paper} (red circles with error bars) and for the galaxy clusters {analysed by} \citet{Capozziello2009} (green circles with error bars).}
		\label{fig:3}
	\end{figure*}

	Figure \ref{fig:3} compares our estimates of $\delta$ and $\xi$ for the eight dSphs.  For $\xi^{-1}$, the dSphs have fully consistent values with mean $\xi^{-1}= 1.2^{+18.6}_{-0.9}$~Mpc, whereas $\delta$ clusters around two values $\overline{\delta}=-0.986\pm0.002$ and $\overline{\delta}=-0.92\pm0.01$;
	these two values disagree by $6\sigma$.  This discrepancy is not consistent with the expectations of $f(R)$ gravity: although $\delta$ and $\xi$ are not universal parameters, they should assume roughly the same values within the same class of objects \citep{Capozziello2011, Capozziello2012}. This tension might be alleviated by improving the model of the dSphs: in our analysis we assume non-rotating spherically symmetric galaxies,  
	a single stellar population, and velocity anisotropy parameter and mass-to-light ratio constant with radius.
	These assumptions clearly are simplistic: they might introduce systematic errors and be responsible for these two wildly different values of $\delta$.
	
	In the literature, the gravitational potential we use here [Eq.\eqref{3e}] was only adopted by \citet{Capozziello2009} to model the mass  profile of a sample of 12 X-ray galaxy clusters. The estimates of $\delta$ and $\xi$ for the X-ray clusters are shown in Fig.~\ref{fig:3}.  Both $\delta$ and $\xi$ appear to depend on the cluster, at odds with the $f(R)$ expectations of consistent values within the same class of objects. 
	The uncertainties on the estimates of $\delta$ and $\xi$ of the clusters are substantially smaller than the uncertainties for the dSphs, where the variation of the parameter values from dSph to dSph appears to be limited. The large fluctuations for the clusters might suggest that, in the cluster analysis, either the random errors are underestimated or some systematic error is present, or both. 
	
\begin{figure*}
		\includegraphics[width=2\columnwidth,keepaspectratio]{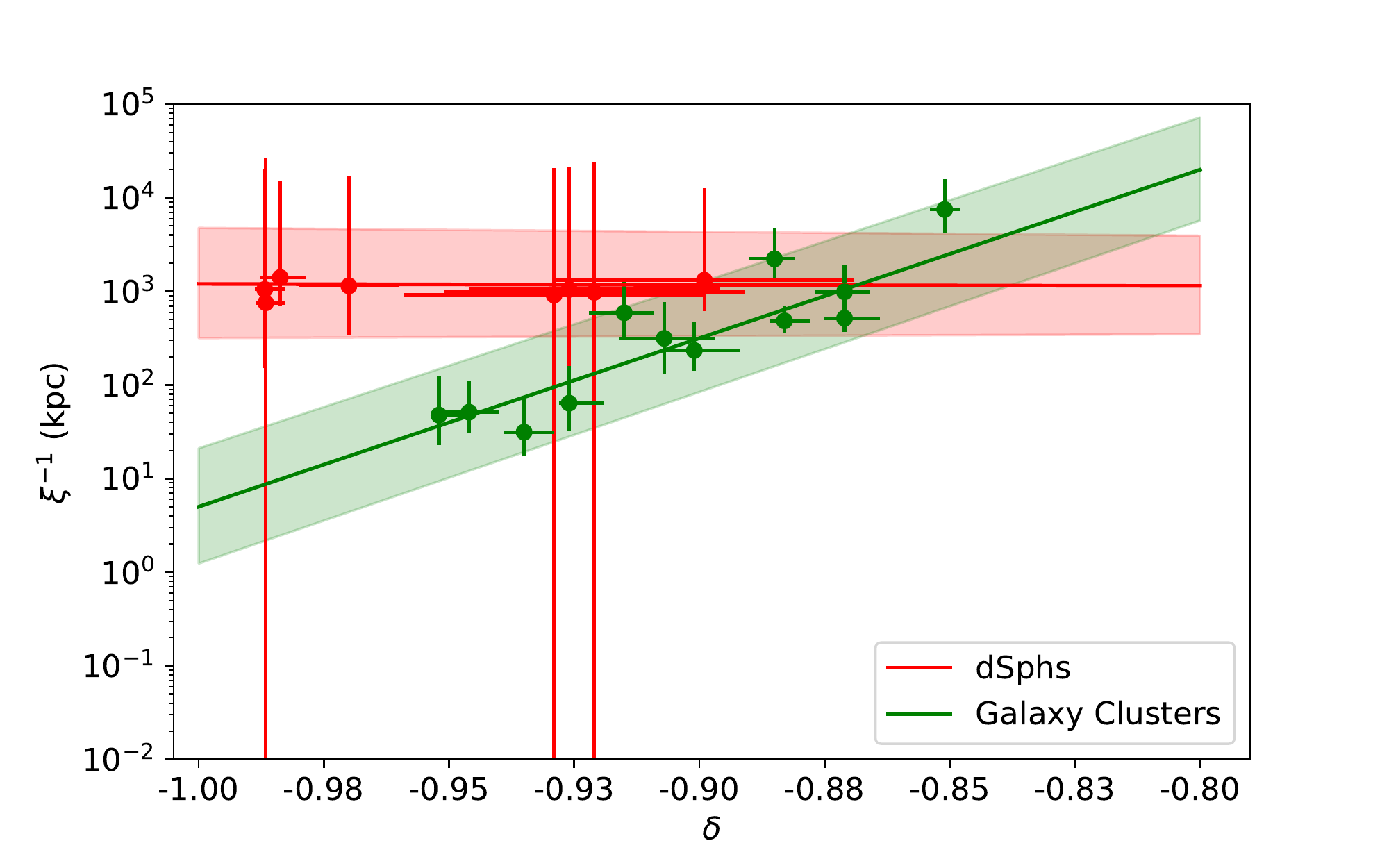}
		\caption{Correlation between the two parameters $\xi^{-1}$ and $\delta$ for dSphs (red solid line) and galaxy clusters (green solid line), respectively.  The solid lines and shaded areas are the least squares fits and their 1-$\sigma$ spreads. }
		\label{fig:4}
	\end{figure*}

	The presence of possible unidentified systematic errors is indeed suggested by the correlation, for the clusters, between the two parameters $\xi$ and $\delta$ shown in Fig.~\ref{fig:4}. The linear correlation coefficient is $\rho=0.90$ with a $p$-value, which is the probability of measuring $\rho$ from uncorrelated quantities, $p=7\times 10^{-5}$. This correlation indicates that the pair $(\delta,\xi)$ depends on the cluster. In principle, this correlation cannot be physically motivated in $f(R)$ gravity, unless we wish to introduce substantial complications in the theory. On the contrary, the values of the dSphs show no correlation ($\rho= 0.14$ and $p=0.75$), supporting the conclusion that systematic errors are less relevant in our dSph analysis than in the cluster analysis. 
	
	\begin{figure*}
		\begin{center}
			\begin{tabular}{c}
				\includegraphics[width=0.99\columnwidth]{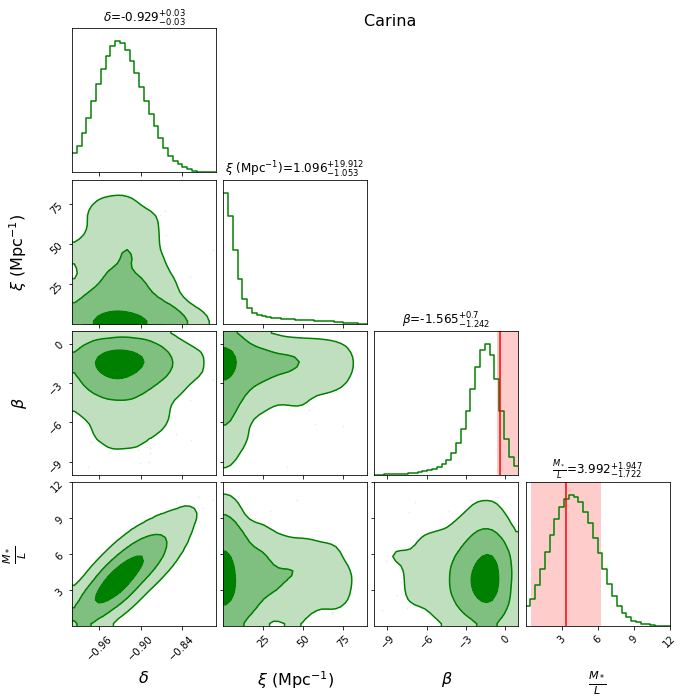}
				\includegraphics[width=0.99\columnwidth]{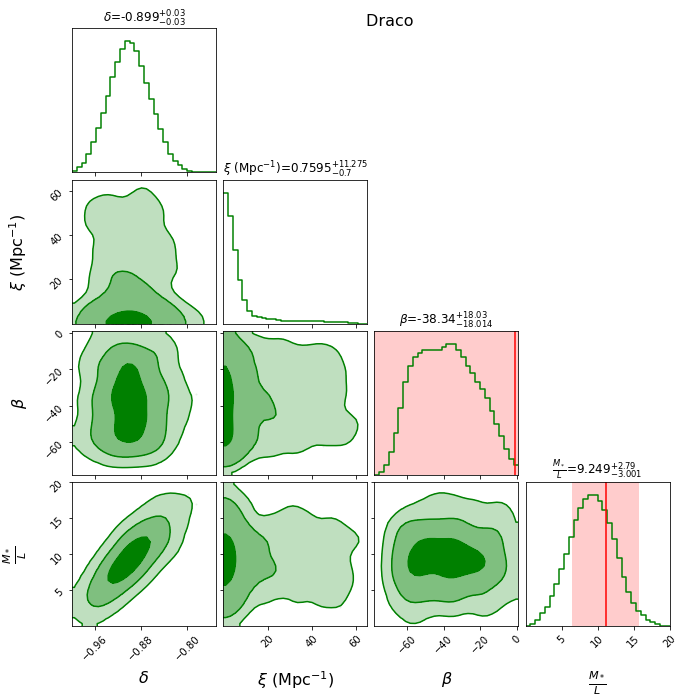}\\
				\includegraphics[width=0.99\columnwidth]{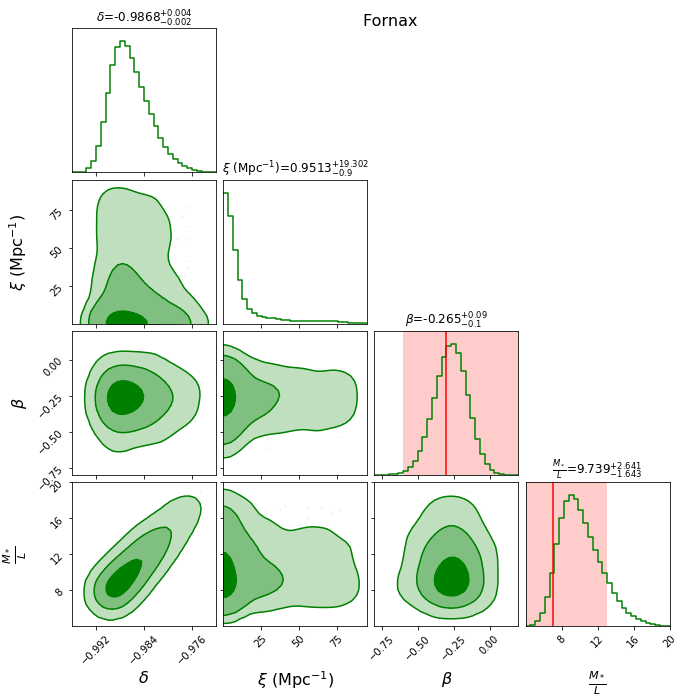}
				\includegraphics[width=0.99\columnwidth]{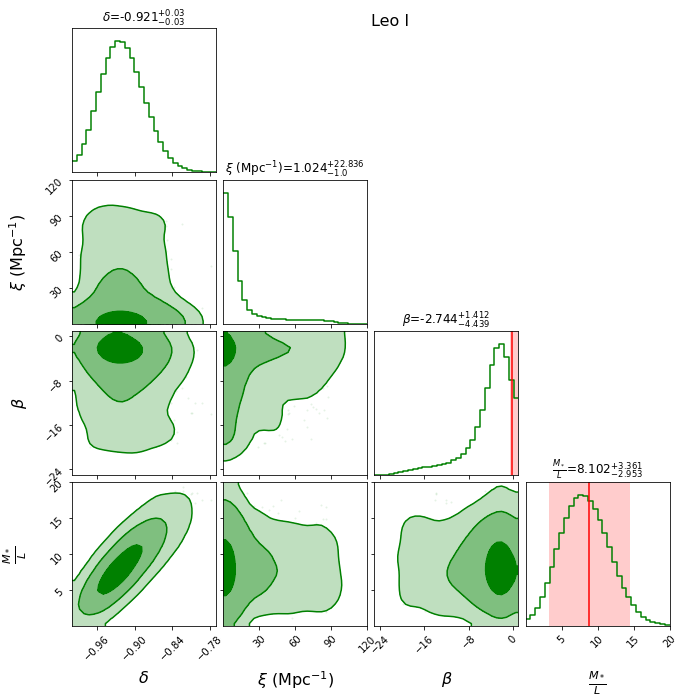}
			\end{tabular}
		\end{center}
		\caption{MCMC posterior distributions of the { set of} parameters $\bm{\theta} = \{\delta, \xi, \beta, M_*/L\}$ for each dSph. The shaded areas with decreasing darkness depict the 68\%, 95\%, and 99\% confidence regions of the posterior distributions, respectively.  The medians of the posterior distributions with their 68\% confidence intervals are reported on top of each column. 
			The red shaded areas in the  panels corresponding to the velocity anisotropy parameter $\beta$ indicate the best fit {values} and the $1\sigma$  {uncertainties} of  $\beta_{\mathrm{NFW}}$ as reported in \citet{Walker2009d} and listed in Column (7) of Table \ref{tab:1}. The red shaded areas in the panels corresponding to the mass-to-light ratio $M_*/L_V$ indicate the values expected from the stellar population synthesis model listed in Column (13) of Table \ref{tab:1}.}
		\label{fig:1}
	\end{figure*}
	
	\begin{figure*}
		\ContinuedFloat
		\captionsetup{list=off,format=cont}
		\begin{center}
			\begin{tabular}{c}
				\includegraphics[width=0.99\columnwidth]{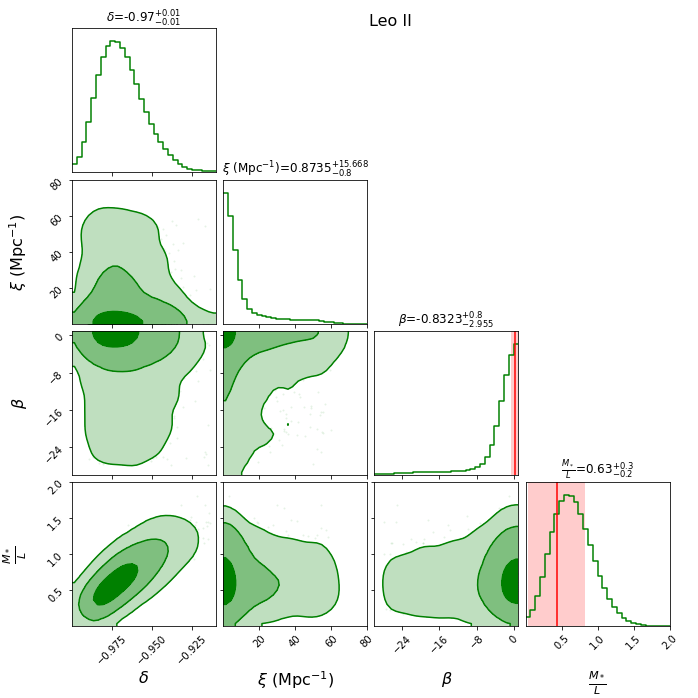}
				\includegraphics[width=0.99\columnwidth]{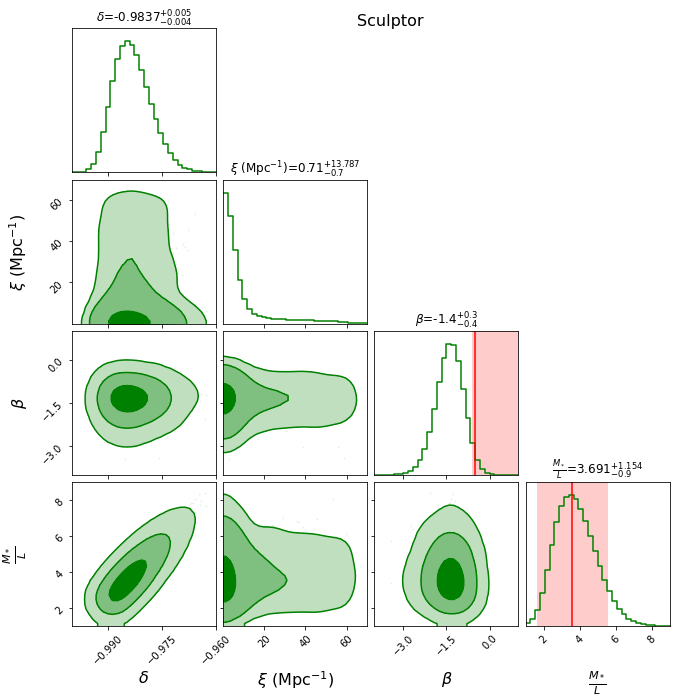}\\
				\includegraphics[width=0.99\columnwidth]{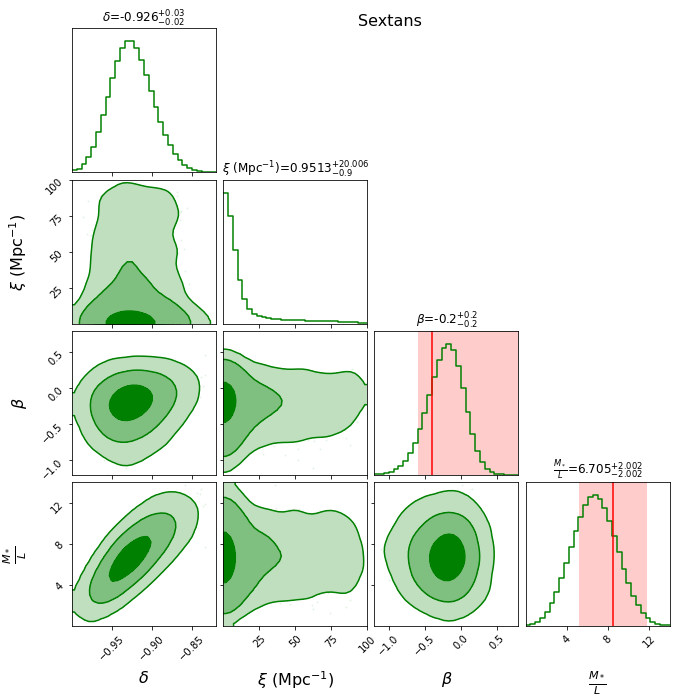}
				\includegraphics[width=0.99\columnwidth]{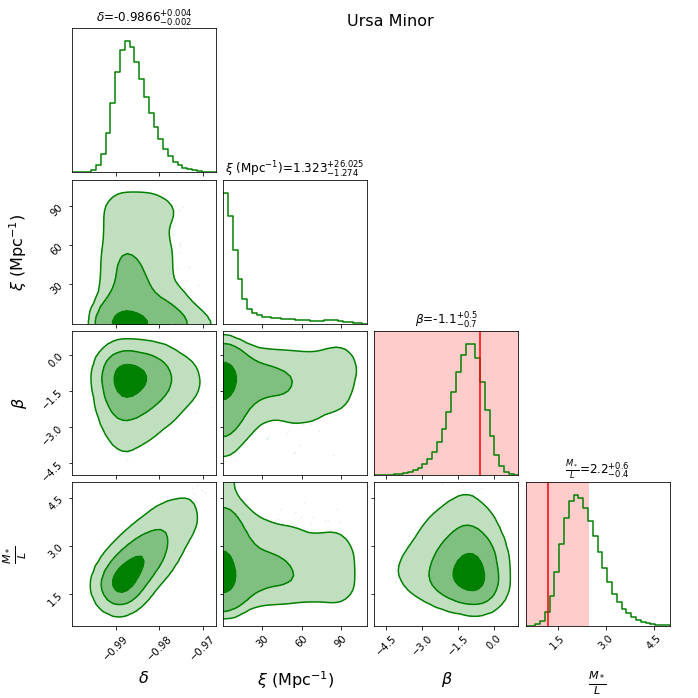}
			\end{tabular}
		\end{center}
		\caption{{\it Continued.}}
	\end{figure*}

	\begin{table*}
		\begin{center}
			\resizebox{14cm}{!}{
				\setlength{\tabcolsep}{4pt}
				\begin{tabular}{lcccc||cc||c}
					\hline
					\hline
					Galaxy & $\delta$ &  $\xi$  & $\beta$ & $M_*/L_V$  & $f'_0$ & $f''_0$   & $\beta_{\mathrm{NFW}}$ \\
					&     & ($\rm{Mpc}^{-1})$  & & &   & ($\rm{Mpc}^{2}$) & \\
					(1)         & (2)       & (3) & (4)  & (5)& (6)  & (7)        & (8) \\
					\hline
					\textbf{Carina} & $-0.93\pm0.03$ & $1^{+20}_{-1}$ 
					& $-1.6^{+0.7}_{-1.3}$ & $4.0^{+1.9}_{-1.7}$ & $0.08^{+0.05}_{-0.03}$ &$0.001^{+0.05}_{-2.77}$  & $-0.4^{+0.6}_{-0.2}$ \\[0.1cm]
					\textbf{Draco} & $-0.90\pm0.03$ & $0.8^{+11.3}_{-0.7}$ & $-38\pm18$ & $9.3^{+2.8}_{-3.0}$ & $0.10\pm0.03$ & $0.0006^{+3.615}_{-0.0006}$ & $-0.6^{+0.8}_{}$ \\[0.1cm]
					\textbf{Fornax} & $-0.987^{+0.004}_{-0.002}$ & $1^{+19}_{-1}$
					& $-0.27^{+0.09}_{-0.1}$ & $9.7^{+2.6}_{-1.6}$ & $0.014^{+0.005}_{-0.002}$ & $0.001^{+0.90}_{-0.001}$ &  $-0.3^{+0.3}_{-0.3}$ \\[0.1cm]
					\textbf{Leo I} & $-0.92\pm0.03$ & $1^{+23}_{-1}$ 
					&  $-2.7^{+1.4}_{-4.4}$ & $8.1^{+3.6}_{-3.0}$ &  $0.08\pm0.03$ & $0.01^{+6.57}_{-0.01}$ & $-0.2^{+0.6}_{-0.3}$ \\[0.1cm]
					\textbf{Leo II}& $-0.97\pm0.01$  & $0.8^{+15.7}_{-0.8}$ & $-0.8^{+0.8}_{-3.0}$ & $0.6^{+0.3}_{-0.2}$ &  $0.03^{+0.02}_{-0.01}$ & $0.002^{+1.68}_{-0.002}$ & $+0.3^{}_{-0.9}$  \\[0.1cm]
					\textbf{Sculptor} & $-0.985^{+0.005}_{-0.004}$ & $0.7^{+13.8}_{-0.7}$ & $-1.4^{+0.3}_{-0.4}$ & $3.7^{+1.2}_{-0.9}$ & $0.016^{+0.005}_{-0.004}$ & $0.00007^{+0.005}_{-0.00007}$ & $-0.5^{+0.5}_{-0.1}$\\[0.1cm]
					\textbf{Sextans} & $-0.93^{+0.03}_{-0.02}$ & $1^{+20}_{-1}$ 
					& $-0.2\pm0.2$ &  $6.7\pm2.0$& $0.07^{+0.03}_{-0.02}$ & $0.01^{+6.61}_{-0.01}$ & $-0.4^{+0.6}_{-0.2}$ \\[0.1cm]
					\textbf{Ursa Minor}& $-0.978^{+0.004}_{-0.003}$ & $1^{+26}_{-1}$ 
					& $-1.1^{+0.5}_{-0.7}$ & $2.2^{+0.6}_{-0.4}$ &  $0.013^{+0.004}_{-0.002}$&  $0.001^{+1.0}_{-0.001}$ & $-0.6^{+0.8}_{}$ \\[0.1cm]
					\hline
				\end{tabular}
			}
		\end{center}
		\caption{MCMC estimates of the { set of} parameters $\bm{\theta} = \{\delta, \xi, \beta, M_*/L_V\}$.  Column (1): dSph name; Columns (2)-(5): median and 1$\sigma$ confidence interval; Columns (6) and (7): Taylor coefficients $f'_0$ and $f''_0${, with}  1$\sigma$ confidence intervals; Columns (8): velocity anisotropy parameter $\beta_{\mathrm{NFW}}$   {under the assumption of} a Navarro-Frank-White model for the dark matter halo in standard gravity \citep{Walker2009d}.}\label{tab:2}
	\end{table*}
	
	\section{Discussion and Conclusions}\label{sec:discussions}
	
	Modifications of GR have been extensively studied to overcome some of the tensions affecting the $\Lambda$CDM model \citep{deMartino2020,Peebles2022}.  $f(R)$ gravity is one of the most tested  theories of gravity on extra-galactic and cosmological scales. Indeed, the extra degrees of freedom that arise in $f(R)$ can easily explain the accelerated expansion of the Universe \citep{Starobinsky2007,WayneHU2007,Amendola2008,Miranda2009,deMartino2015,Lazkoz_2018,Calza2018}, whereas $f(R)$ reduces to GR on  the Solar System  scale \citep{DeFelice2010}. On the intermediate scale of galaxies, the ability of $f(R)$ gravity to describe the dynamics of cosmic structure without the aid of dark matter, as in the standard model, is substantially less investigated. 
	
	In the weak-field limit, $f(R)$ introduces 
	a modification to the Newtonian gravitational potential that is parameterized by the two parameters $\delta$ and $\xi$ 
	\citep{Capozziello2009,Banerjee2017}.  
	Here, we contributed to the quest of the ability of these two parameters to play the role of dark matter by assessing whether $f(R)$ gravity can describe the dynamics of eight dSphs. We modelled 
	the observed velocity dispersion profiles of the dSph stars with the Jeans equation (Sect.~ \ref{sec:mcmc}). 
	Our results are based on a set of simplifying assumptions: a non-rotating spherically symmetric galaxy,  a single
	stellar population, a mass-to-light ratio and a velocity anisotropy parameter $\beta$ constant with radius. 
	
	Our analysis returns a length scale of the Yukawa potential, $\xi^{-1}= 1.2^{+18.6}_{-0.9}$~Mpc for all the galaxies. This result is reassuring for two reasons: (1) a single value of $\xi^{-1}$ is valid for the entire dSph sample, as expected in $f(R)$ gravity within the same class of objects; (2) the value of $\xi^{-1}$ guarantees that the extra-degrees of freedom do not play any relevant role in the dynamics of self-gravitating systems on scale below $\sim 1$~Mpc, where the effects of the accelerated expansion of the Universe or the effects of a fifth force are lacking. 
	Our result confirms previous analyses. For instance, \citet{DeMartino2016} found $\xi^{-1}\sim 1$~Mpc by fitting the SZ temperature anisotropy profile of the Coma cluster. \citet{Capozziello2009} and \citet{Napolitano2012} also found similar values of $\xi^{-1}$ 
	by modelling the mass profile of 12 X-ray galaxy clusters and  the stellar kinematics of elliptical galaxies, respectively. Actually, the estimated values of $\xi^{-1}$ in \citet{Capozziello2009} range from 100 kpc to 10 Mpc depending on the systems. However, these fluctuations of $\xi^{-1}$  may be due to a simplistic modeling based on the assumption of a phenomenological X-ray gas density whose parameters are not fitted together with the $f(R)$-gravity parameters. 
	
	On the other hand, our result on $\delta$ suggests a possible tension for $f(R)$ gravity. The parameter $\delta$ controls the intensity of the gravitational field. In Eq. (\ref{3e}) for the modified gravitational potential of an extended mass distribution, the term $(1+\delta)^{-1}$  multiplies the mass density: $\delta$ can thus mimic an increasing  dark matter content by assuming values increasingly close to $-1$. 
	
	The values of $\delta$ we find for the dSph sample have a bimodal distribution peaking around two values $\overline{\delta}=-0.986\pm0.002$ and $\overline{\delta}=-0.92\pm0.01$. These values differ by 6-$\sigma$. Therefore, unlike $\xi$, $\delta$ does not assume a single value which is valid for all the dSphs in the sample. Our analysis rather suggests that different values of $\delta$ are required for different objects within the same class, at odds with the expectations from $f(R)$ gravity.
	
	Different values of $\delta$ are expected in different classes of objects but not within the same class of objects. For example, for gravitational systems where dark matter is not required in standard gravity, like stellar systems, $\delta$ is expected to be sufficiently close to zero to guarantee negligible departures, if any, from the standard Newtonian dynamics. Indeed, \citet{deMartino2021} showed that reproducing the orbital motion of the S2 star around the supermassive black hole in the centre of the Galaxy requires $\delta=-0.01_{-0.14}^{+0.61}$, which is compatible with zero at 1-$\sigma$.  
	On the contrary, for systems dominated by dark matter in the standard model, like clusters of galaxies, $\delta$ is close to $-1$. For example,  \citet{Capozziello2009}  found  $\delta$ in the range $[-0.95,-0.85]$ for the X-ray clusters mentioned above. 
	The disagreement we find for the two values of $\delta$ for the dSphs suggests a severe tension for $f(R)$ gravity. 
	
	{ In principle, more sophisticated extended theories of gravity might properly describe 
		the kinematics of dwarf galaxies  and reduce, or completely eliminate,  the tension we find here. For example, a 
		scalar-tensor $f(R,\phi)$ theory, a fourth-order theory of gravity non-minimally coupled with a massive scalar field $\phi$,
		generates Yukawa-like corrections to the Newtonian gravitational potential in the weak-filed limit similarly to $f(R)$ gravity. The correction terms are related to the extra degrees of freedom and  generate 
		gravitational forces that 
		mimic the gravitational field of the dark matter halo and accommodate the rotation curves of spiral galaxies \citep{2013PhRvD..87f4002S}. It is likely, but still to be proved, that similar correction terms can also describe the dynamics of dwarf galaxies.  
		
		Alternatively, improved models of the dwarf galaxies, where our simplifying assumptions are dropped, may be sufficient to remove the tension for $f(R)$ gravity.
		Further insights will come from observations:} a  future {\it Theia}-like astrometric mission \citep{Malbet2016,Theia2017,Malbet2019,Malbet2021} 
	will provide high precision measures of the proper motion of the dSph stars.
	Similarly to what we expect for the core-cusp problem \citep{deMartino2022}, these additional information will decrease the current uncertainties on the estimate of $\xi^{-1}$, and either confirm or remove the tension on $\delta$ we find here.
	
	\section*{Acknowledgements}
	
	We acknowledge partial support from the INFN grant InDark and the Italian Ministry of Education, University and Research (MIUR) under the Departments of Excellence grant L.232/2016. 
	IDM acknowledges support from Ayuda  IJCI2018-036198-I  funded by  MCIN/AEI/  10.13039/501100011033  and  FSE  “El FSE  invierte  en  tu  futuro”  o  “Financiado  por  la  Unión  Europea   “NextGenerationEU”/PRTR. 
	IDM is also supported by the project PID2021-122938NB-I00  funded by the Spanish "Ministerio de Ciencia e Innovación" and FEDER “A way of making Europe", and by the project SA096P20 Junta de Castilla y León. 
	This research has made use of NASA’s Astrophysics Data System Bibliographic Services.
	
	\section*{Data Availability Statement}
	Data are publicly available in \citep{Walker2007,Walker2009a,Walker2009b,Walker2009c,Walker2009d}.
	
	
	
	
	\bibliographystyle{mnras}
	\bibliography{ms_refs} 

\begin{thebibliography}{}
\makeatletter
\relax
\def\mn@urlcharsother{\let\do\@makeother \do\$\do\&\do\#\do\^\do\_\do\%\do\~}
\def\mn@doi{\begingroup\mn@urlcharsother \@ifnextchar [ {\mn@doi@}
  {\mn@doi@[]}}
\def\mn@doi@[#1]#2{\def\@tempa{#1}\ifx\@tempa\@empty \href
  {http://dx.doi.org/#2} {doi:#2}\else \href {http://dx.doi.org/#2} {#1}\fi
  \endgroup}
\def\mn@eprint#1#2{\mn@eprint@#1:#2::\@nil}
\def\mn@eprint@arXiv#1{\href {http://arxiv.org/abs/#1} {{\tt arXiv:#1}}}
\def\mn@eprint@dblp#1{\href {http://dblp.uni-trier.de/rec/bibtex/#1.xml}
  {dblp:#1}}
\def\mn@eprint@#1:#2:#3:#4\@nil{\def\@tempa {#1}\def\@tempb {#2}\def\@tempc
  {#3}\ifx \@tempc \@empty \let \@tempc \@tempb \let \@tempb \@tempa \fi \ifx
  \@tempb \@empty \def\@tempb {arXiv}\fi \@ifundefined
  {mn@eprint@\@tempb}{\@tempb:\@tempc}{\expandafter \expandafter \csname
  mn@eprint@\@tempb\endcsname \expandafter{\@tempc}}}

\bibitem[\protect\citeauthoryear{{Amendola} \& {Tsujikawa}}{{Amendola} \&
  {Tsujikawa}}{2008}]{Amendola2008}
{Amendola} L.,  {Tsujikawa} S.,  2008, Physics Letters B, 660, 125

\bibitem[\protect\citeauthoryear{{Banerjee}, {Shankar}  \& {Singh}}{{Banerjee}
  et~al.}{2017}]{Banerjee2017}
{Banerjee} S.,  {Shankar} S.,   {Singh} T.~P.,  2017, \mn@doi [\jcap]
  {10.1088/1475-7516/2017/10/004}, \href
  {https://ui.adsabs.harvard.edu/abs/2017JCAP...10..004B} {2017, 004}

\bibitem[\protect\citeauthoryear{{Bell} \& {de Jong}}{{Bell} \& {de
  Jong}}{2001}]{Bell2001}
{Bell} E.~F.,  {de Jong} R.~S.,  2001, \mn@doi [\apj] {10.1086/319728}, \href
  {https://ui.adsabs.harvard.edu/abs/2001ApJ...550..212B} {550, 212}

\bibitem[\protect\citeauthoryear{{Bellazzini}, {Gennari}, {Ferraro}  \&
  {Sollima}}{{Bellazzini} et~al.}{2004}]{Bellazzini2004}
{Bellazzini} M.,  {Gennari} N.,  {Ferraro} F.~R.,   {Sollima} A.,  2004,
  \mn@doi [\mnras] {10.1111/j.1365-2966.2004.08226.x}, \href
  {https://ui.adsabs.harvard.edu/abs/2004MNRAS.354..708B} {354, 708}

\bibitem[\protect\citeauthoryear{{Bellazzini}, {Gennari}  \&
  {Ferraro}}{{Bellazzini} et~al.}{2005}]{Bellazzini2005}
{Bellazzini} M.,  {Gennari} N.,   {Ferraro} F.~R.,  2005, \mn@doi [\mnras]
  {10.1111/j.1365-2966.2005.09027.x}, \href
  {https://ui.adsabs.harvard.edu/abs/2005MNRAS.360..185B} {360, 185}

\bibitem[\protect\citeauthoryear{{Binney} \& {Tremaine}}{{Binney} \&
  {Tremaine}}{2008}]{Binney2008}
{Binney} J.,  {Tremaine} S.,  2008, {Galactic Dynamics: Second Edition}

\bibitem[\protect\citeauthoryear{{Bonanos}, {Stanek}, {Szentgyorgyi},
  {Sasselov}  \& {Bakos}}{{Bonanos} et~al.}{2004}]{Bonanos2004}
{Bonanos} A.~Z.,  {Stanek} K.~Z.,  {Szentgyorgyi} A.~H.,  {Sasselov} D.~D.,
  {Bakos} G.~{\'A}.,  2004, \mn@doi [\aj] {10.1086/381073}, \href
  {https://ui.adsabs.harvard.edu/abs/2004AJ....127..861B} {127, 861}

\bibitem[\protect\citeauthoryear{{Boylan-Kolchin}, {Bullock}  \&
  {Kaplinghat}}{{Boylan-Kolchin} et~al.}{2011}]{Boylan_Kolchin2011}
{Boylan-Kolchin} M.,  {Bullock} J.~S.,   {Kaplinghat} M.,  2011, \mn@doi
  [\mnras] {10.1111/j.1745-3933.2011.01074.x}, \href
  {https://ui.adsabs.harvard.edu/abs/2011MNRAS.415L..40B} {415, L40}

\bibitem[\protect\citeauthoryear{{Bullock} \& {Boylan-Kolchin}}{{Bullock} \&
  {Boylan-Kolchin}}{2017}]{Bullock2017}
{Bullock} J.~S.,  {Boylan-Kolchin} M.,  2017, \mn@doi [\araa]
  {10.1146/annurev-astro-091916-055313}, \href
  {https://ui.adsabs.harvard.edu/abs/2017ARA&A..55..343B} {55, 343}

\bibitem[\protect\citeauthoryear{{Calz{\`a}}, {Rinaldi}  \&
  {Sebastiani}}{{Calz{\`a}} et~al.}{2018}]{Calza2018}
{Calz{\`a}} M.,  {Rinaldi} M.,   {Sebastiani} L.,  2018, \mn@doi [European
  Physical Journal C] {10.1140/epjc/s10052-018-5681-8}, \href
  {https://ui.adsabs.harvard.edu/abs/2018EPJC...78..178C} {78, 178}

\bibitem[\protect\citeauthoryear{{Capolupo}}{{Capolupo}}{2017}]{Capolupo2017}
{Capolupo} A.,  2017, \mn@doi [Galaxies] {10.3390/galaxies5040098}, \href
  {https://ui.adsabs.harvard.edu/abs/2017Galax...5...98C} {5, 98}

\bibitem[\protect\citeauthoryear{{Capozziello} \& {De Laurentis}}{{Capozziello}
  \& {De Laurentis}}{2012}]{Capozziello2012}
{Capozziello} S.,  {De Laurentis} M.,  2012, \mn@doi [Annalen der Physik]
  {10.1002/andp.201200109}, \href
  {https://ui.adsabs.harvard.edu/abs/2012AnP...524..545C} {524, 545}

\bibitem[\protect\citeauthoryear{{Capozziello} \& {Tsujikawa}}{{Capozziello} \&
  {Tsujikawa}}{2008}]{Capozziello2008}
{Capozziello} S.,  {Tsujikawa} S.,  2008, \mn@doi [\prd]
  {10.1103/PhysRevD.77.107501}, \href
  {https://ui.adsabs.harvard.edu/abs/2008PhRvD..77j7501C} {77, 107501}

\bibitem[\protect\citeauthoryear{{Capozziello} \& {de Laurentis}}{{Capozziello}
  \& {de Laurentis}}{2011}]{Capozziello2011}
{Capozziello} S.,  {de Laurentis} M.,  2011, \mn@doi [\physrep]
  {10.1016/j.physrep.2011.09.003}, \href
  {https://ui.adsabs.harvard.edu/abs/2011PhR...509..167C} {509, 167}

\bibitem[\protect\citeauthoryear{{Capozziello}, {Stabile}  \&
  {Troisi}}{{Capozziello} et~al.}{2007}]{Capozziello2007}
{Capozziello} S.,  {Stabile} A.,   {Troisi} A.,  2007, \mn@doi [\prd]
  {10.1103/PhysRevD.76.104019}, \href
  {https://ui.adsabs.harvard.edu/abs/2007PhRvD..76j4019C} {76, 104019}

\bibitem[\protect\citeauthoryear{{Capozziello}, {de Filippis}  \&
  {Salzano}}{{Capozziello} et~al.}{2009}]{Capozziello2009}
{Capozziello} S.,  {de Filippis} E.,   {Salzano} V.,  2009, \mn@doi [\mnras]
  {10.1111/j.1365-2966.2008.14382.x}, \href
  {https://ui.adsabs.harvard.edu/abs/2009MNRAS.394..947C} {394, 947}

\bibitem[\protect\citeauthoryear{{Carrera}, {Aparicio}, {Mart{\'\i}nez-Delgado}
   \& {Alonso-Garc{\'\i}a}}{{Carrera} et~al.}{2002}]{Carrera2002}
{Carrera} R.,  {Aparicio} A.,  {Mart{\'\i}nez-Delgado} D.,
  {Alonso-Garc{\'\i}a} J.,  2002, \mn@doi [\aj] {10.1086/340702}, \href
  {https://ui.adsabs.harvard.edu/abs/2002AJ....123.3199C} {123, 3199}

\bibitem[\protect\citeauthoryear{{Cautun} et~al.,}{{Cautun}
  et~al.}{2020}]{Cautun2020}
{Cautun} M.,  et~al., 2020, \mn@doi [\mnras] {10.1093/mnras/staa1017}, \href
  {https://ui.adsabs.harvard.edu/abs/2020MNRAS.494.4291C} {494, 4291}

\bibitem[\protect\citeauthoryear{{Cognola}, {Elizalde}, {Nojiri}, {Odintsov},
  {Sebastiani}  \& {Zerbini}}{{Cognola} et~al.}{2008}]{Cognola2008}
{Cognola} G.,  {Elizalde} E.,  {Nojiri} S.,  {Odintsov} S.~D.,  {Sebastiani}
  L.,   {Zerbini} S.,  2008, \mn@doi [\prd] {10.1103/PhysRevD.77.046009}, \href
  {https://ui.adsabs.harvard.edu/abs/2008PhRvD..77d6009C} {77, 046009}

\bibitem[\protect\citeauthoryear{{De Felice} \& {Tsujikawa}}{{De Felice} \&
  {Tsujikawa}}{2010}]{DeFelice2010}
{De Felice} A.,  {Tsujikawa} S.,  2010, \mn@doi [Living Reviews in Relativity]
  {10.12942/lrr-2010-3}, \href
  {https://ui.adsabs.harvard.edu/abs/2010LRR....13....3D} {13, 3}

\bibitem[\protect\citeauthoryear{{De Laurentis} \& {De Martino}}{{De Laurentis}
  \& {De Martino}}{2013}]{DeLaurentis2013}
{De Laurentis} M.,  {De Martino} I.,  2013, \mn@doi [\mnras]
  {10.1093/mnras/stt216}, \href
  {https://ui.adsabs.harvard.edu/abs/2013MNRAS.431..741D} {431, 741}

\bibitem[\protect\citeauthoryear{{De Laurentis}, {De Martino}  \& {Lazkoz}}{{De
  Laurentis} et~al.}{2018}]{DeLaurentis2018}
{De Laurentis} M.,  {De Martino} I.,   {Lazkoz} R.,  2018, \mn@doi [\prd]
  {10.1103/PhysRevD.97.104068}, \href
  {https://ui.adsabs.harvard.edu/abs/2018PhRvD..97j4068D} {97, 104068}

\bibitem[\protect\citeauthoryear{{De Martino}}{{De
  Martino}}{2016}]{DeMartino2016}
{De Martino} I.,  2016, \mn@doi [\prd] {10.1103/PhysRevD.93.124043}, \href
  {https://ui.adsabs.harvard.edu/abs/2016PhRvD..93l4043D} {93, 124043}

\bibitem[\protect\citeauthoryear{{De Martino}, {De Laurentis},
  {Atrio-Barandela}  \& {Capozziello}}{{De Martino}
  et~al.}{2014}]{DeMartino2014}
{De Martino} I.,  {De Laurentis} M.,  {Atrio-Barandela} F.,   {Capozziello} S.,
   2014, \mn@doi [\mnras] {10.1093/mnras/stu903}, \href
  {https://ui.adsabs.harvard.edu/abs/2014MNRAS.442..921D} {442, 921}

\bibitem[\protect\citeauthoryear{{De Martino}, {De Laurentis}  \&
  {Capozziello}}{{De Martino} et~al.}{2015}]{deMartino2015}
{De Martino} I.,  {De Laurentis} M.,   {Capozziello} S.,  2015, \mn@doi
  [Universe] {10.3390/universe1020123}, \href
  {https://ui.adsabs.harvard.edu/abs/2015Univ....1..123D} {1, 123}

\bibitem[\protect\citeauthoryear{{De Martino}, {Lazkoz}  \& {De Laurentis}}{{De
  Martino} et~al.}{2018}]{DeMartino2018}
{De Martino} I.,  {Lazkoz} R.,   {De Laurentis} M.,  2018, \mn@doi [\prd]
  {10.1103/PhysRevD.97.104067}, \href
  {https://ui.adsabs.harvard.edu/abs/2018PhRvD..97j4067D} {97, 104067}

\bibitem[\protect\citeauthoryear{{De Martino}, {Chakrabarty}, {Cesare},
  {Gallo}, {Ostorero}  \& {Diaferio}}{{De Martino}
  et~al.}{2020}]{deMartino2020}
{De Martino} I.,  {Chakrabarty} S.~S.,  {Cesare} V.,  {Gallo} A.,  {Ostorero}
  L.,   {Diaferio} A.,  2020, \mn@doi [Universe] {10.3390/universe6080107},
  \href {https://ui.adsabs.harvard.edu/abs/2020Univ....6..107D} {6, 107}

\bibitem[\protect\citeauthoryear{{De Martino}, {della Monica}  \& {De
  Laurentis}}{{De Martino} et~al.}{2021}]{deMartino2021}
{De Martino} I.,  {della Monica} R.,   {De Laurentis} M.,  2021, \mn@doi [\prd]
  {10.1103/PhysRevD.104.L101502}, \href
  {https://ui.adsabs.harvard.edu/abs/2021PhRvD.104j1502D} {104, L101502}

\bibitem[\protect\citeauthoryear{De~Martino, Diaferio  \& Ostorero}{De~Martino
  et~al.}{2022}]{deMartino2022}
De~Martino I.,  Diaferio A.,   Ostorero L.,  2022, \mn@doi [Monthly Notices of
  the Royal Astronomical Society] {10.1093/mnras/stac2336}

\bibitem[\protect\citeauthoryear{{Del Popolo} \& {Le Delliou}}{{Del Popolo} \&
  {Le Delliou}}{2017}]{DelPopolo2017}
{Del Popolo} A.,  {Le Delliou} M.,  2017, \mn@doi [Galaxies]
  {10.3390/galaxies5010017}, \href
  {https://ui.adsabs.harvard.edu/abs/2017Galax...5...17D} {5, 17}

\bibitem[\protect\citeauthoryear{{Di Valentino} et~al.,}{{Di Valentino}
  et~al.}{2021}]{DiValentino2021}
{Di Valentino} E.,  et~al., 2021, \mn@doi [Classical and Quantum Gravity]
  {10.1088/1361-6382/ac086d}, \href
  {https://ui.adsabs.harvard.edu/abs/2021CQGra..38o3001D} {38, 153001}

\bibitem[\protect\citeauthoryear{{Foreman-Mackey}, {Hogg}, {Lang}  \&
  {Goodman}}{{Foreman-Mackey} et~al.}{2013}]{emcee}
{Foreman-Mackey} D.,  {Hogg} D.~W.,  {Lang} D.,   {Goodman} J.,  2013, \mn@doi
  [Publications of the Astronomical Society of the Pacific] {10.1086/670067},
  \href {https://ui.adsabs.harvard.edu/abs/2013PASP..125..306F} {125, 306}

\bibitem[\protect\citeauthoryear{{Fritz}, {Battaglia}, {Pawlowski},
  {Kallivayalil}, {van der Marel}, {Sohn}, {Brook}  \& {Besla}}{{Fritz}
  et~al.}{2018}]{Fritz2018}
{Fritz} T.~K.,  {Battaglia} G.,  {Pawlowski} M.~S.,  {Kallivayalil} N.,  {van
  der Marel} R.,  {Sohn} S.~T.,  {Brook} C.,   {Besla} G.,  2018, \mn@doi
  [\aap] {10.1051/0004-6361/201833343}, \href
  {https://ui.adsabs.harvard.edu/abs/2018A&A...619A.103F} {619, A103}

\bibitem[\protect\citeauthoryear{{Hu} \& {Sawicki}}{{Hu} \&
  {Sawicki}}{2007}]{WayneHU2007}
{Hu} W.,  {Sawicki} I.,  2007, Phys. Rev. D, 76, 064004

\bibitem[\protect\citeauthoryear{{Irwin} \& {Hatzidimitriou}}{{Irwin} \&
  {Hatzidimitriou}}{1995}]{Irwin1995}
{Irwin} M.,  {Hatzidimitriou} D.,  1995, \mn@doi [\mnras]
  {10.1093/mnras/277.4.1354}, \href
  {https://ui.adsabs.harvard.edu/abs/1995MNRAS.277.1354I} {277, 1354}

\bibitem[\protect\citeauthoryear{{Kennicutt}}{{Kennicutt}}{1983}]{Kennicutt1983}
{Kennicutt} R.~C. J.,  1983, \mn@doi [\apj] {10.1086/161261}, \href
  {https://ui.adsabs.harvard.edu/abs/1983ApJ...272...54K} {272, 54}

\bibitem[\protect\citeauthoryear{{Khoury} \& {Weltman}}{{Khoury} \&
  {Weltman}}{2004}]{Khoury2004}
{Khoury} J.,  {Weltman} A.,  2004, \mn@doi [\prd] {10.1103/PhysRevD.69.044026},
  \href {https://ui.adsabs.harvard.edu/abs/2004PhRvD..69d4026K} {69, 044026}

\bibitem[\protect\citeauthoryear{{Kinemuchi}, {Harris}, {Smith}, {Silbermann},
  {Snyder}, {La Cluyz{\'e}}  \& {Clark}}{{Kinemuchi}
  et~al.}{2008}]{Kinemuchi2008}
{Kinemuchi} K.,  {Harris} H.~C.,  {Smith} H.~A.,  {Silbermann} N.~A.,  {Snyder}
  L.~A.,  {La Cluyz{\'e}} A.~P.,   {Clark} C.~L.,  2008, \mn@doi [\aj]
  {10.1088/0004-6256/136/5/1921}, \href
  {https://ui.adsabs.harvard.edu/abs/2008AJ....136.1921K} {136, 1921}

\bibitem[\protect\citeauthoryear{{King}}{{King}}{1962}]{King1962}
{King} I.,  1962, \mn@doi [\aj] {10.1086/108756}, \href
  {https://ui.adsabs.harvard.edu/abs/1962AJ.....67..471K} {67, 471}

\bibitem[\protect\citeauthoryear{{Koch}, {Kleyna}, {Wilkinson}, {Grebel},
  {Gilmore}, {Evans}, {Wyse}  \& {Harbeck}}{{Koch} et~al.}{2007}]{Koch2007}
{Koch} A.,  {Kleyna} J.~T.,  {Wilkinson} M.~I.,  {Grebel} E.~K.,  {Gilmore}
  G.~F.,  {Evans} N.~W.,  {Wyse} R. F.~G.,   {Harbeck} D.~R.,  2007, \mn@doi
  [\aj] {10.1086/519380}, \href
  {https://ui.adsabs.harvard.edu/abs/2007AJ....134..566K} {134, 566}

\bibitem[\protect\citeauthoryear{Lazkoz, Ortiz-Baños  \& Salzano}{Lazkoz
  et~al.}{2018}]{Lazkoz_2018}
Lazkoz R.,  Ortiz-Baños M.,   Salzano V.,  2018, \mn@doi [The European
  Physical Journal C] {10.1140/epjc/s10052-018-5711-6}, 78

\bibitem[\protect\citeauthoryear{{Lee}, {Yuk}, {Park}, {Harris}  \&
  {Zaritsky}}{{Lee} et~al.}{2009}]{Lee2009}
{Lee} M.~G.,  {Yuk} I.-S.,  {Park} H.~S.,  {Harris} J.,   {Zaritsky} D.,  2009,
  \mn@doi [\apj] {10.1088/0004-637X/703/1/692}, \href
  {https://ui.adsabs.harvard.edu/abs/2009ApJ...703..692L} {703, 692}

\bibitem[\protect\citeauthoryear{{L{\'e}pine}, {Koch}, {Rich}  \&
  {Kuijken}}{{L{\'e}pine} et~al.}{2011}]{Lepine2011}
{L{\'e}pine} S.,  {Koch} A.,  {Rich} R.~M.,   {Kuijken} K.,  2011, \mn@doi
  [\apj] {10.1088/0004-637X/741/2/100}, \href
  {https://ui.adsabs.harvard.edu/abs/2011ApJ...741..100L} {741, 100}

\bibitem[\protect\citeauthoryear{{Li} \& {Barrow}}{{Li} \&
  {Barrow}}{2007}]{Li2007}
{Li} B.,  {Barrow} J.,  2007, Phys. Rev. D, 75, 084010

\bibitem[\protect\citeauthoryear{{{\L}okas} \& {Mamon}}{{{\L}okas} \&
  {Mamon}}{2003}]{Lokas2003}
{{\L}okas} E.~L.,  {Mamon} G.~A.,  2003, \mn@doi [\mnras]
  {10.1046/j.1365-8711.2003.06684.x}, \href
  {https://ui.adsabs.harvard.edu/abs/2003MNRAS.343..401L} {343, 401}

\bibitem[\protect\citeauthoryear{{Malbet} et~al.,}{{Malbet}
  et~al.}{2016}]{Malbet2016}
{Malbet} F.,  et~al., 2016, in {MacEwen} H.~A.,  {Fazio} G.~G.,  {Lystrup} M.,
  {Batalha} N.,  {Siegler} N.,   {Tong} E.~C.,  eds,  Society of Photo-Optical
  Instrumentation Engineers (SPIE) Conference Series Vol. 9904, Space
  Telescopes and Instrumentation 2016: Optical, Infrared, and Millimeter Wave.
  p. 99042F, \mn@doi{10.1117/12.2234425}

\bibitem[\protect\citeauthoryear{{Malbet} et~al.,}{{Malbet}
  et~al.}{2019}]{Malbet2019}
{Malbet} F.,  et~al., 2019, arXiv e-prints, \href
  {https://ui.adsabs.harvard.edu/abs/2019arXiv191008028M} {p. arXiv:1910.08028}

\bibitem[\protect\citeauthoryear{Malbet et~al.,}{Malbet
  et~al.}{2021}]{Malbet2021}
Malbet F.,  et~al., 2021, \mn@doi [{Experimental Astronomy}]
  {10.1007/s10686-021-09781-1}, 51, 845

\bibitem[\protect\citeauthoryear{{Mamon} \& {Bou{\'e}}}{{Mamon} \&
  {Bou{\'e}}}{2010}]{Mamon2010}
{Mamon} G.~A.,  {Bou{\'e}} G.,  2010, \mn@doi [\mnras]
  {10.1111/j.1365-2966.2009.15817.x}, \href
  {https://ui.adsabs.harvard.edu/abs/2010MNRAS.401.2433M} {401, 2433}

\bibitem[\protect\citeauthoryear{{Mamon} \& {{\L}okas}}{{Mamon} \&
  {{\L}okas}}{2005}]{Mamon2005}
{Mamon} G.~A.,  {{\L}okas} E.~L.,  2005, \mn@doi [\mnras]
  {10.1111/j.1365-2966.2005.09400.x}, \href
  {https://ui.adsabs.harvard.edu/abs/2005MNRAS.363..705M} {363, 705}

\bibitem[\protect\citeauthoryear{{Martin}, {de Jong}  \& {Rix}}{{Martin}
  et~al.}{2008}]{Martin2008}
{Martin} N.~F.,  {de Jong} J. T.~A.,   {Rix} H.-W.,  2008, \mn@doi [\apj]
  {10.1086/590336}, \href
  {https://ui.adsabs.harvard.edu/abs/2008ApJ...684.1075M} {684, 1075}

\bibitem[\protect\citeauthoryear{{Mateo}}{{Mateo}}{1998}]{Mateo1998}
{Mateo} M.~L.,  1998, \mn@doi [\araa] {10.1146/annurev.astro.36.1.435}, \href
  {https://ui.adsabs.harvard.edu/abs/1998ARA&A..36..435M} {36, 435}

\bibitem[\protect\citeauthoryear{{Mateo}, {Olszewski}  \& {Walker}}{{Mateo}
  et~al.}{2008}]{Mateo2008}
{Mateo} M.,  {Olszewski} E.~W.,   {Walker} M.~G.,  2008, \mn@doi [\apj]
  {10.1086/522326}, \href
  {https://ui.adsabs.harvard.edu/abs/2008ApJ...675..201M} {675, 201}

\bibitem[\protect\citeauthoryear{{Miranda}, {Jor{\'a}s}, {Waga}  \&
  {Quartin}}{{Miranda} et~al.}{2009}]{Miranda2009}
{Miranda} V.,  {Jor{\'a}s} S.~E.,  {Waga} I.,   {Quartin} M.,  2009, Physical
  Review Letters, 102, 221101

\bibitem[\protect\citeauthoryear{{Mu{\~n}oz}, {Majewski}  \&
  {Johnston}}{{Mu{\~n}oz} et~al.}{2008}]{Munoz2008}
{Mu{\~n}oz} R.~R.,  {Majewski} S.~R.,   {Johnston} K.~V.,  2008, \mn@doi [\apj]
  {10.1086/587125}, \href
  {https://ui.adsabs.harvard.edu/abs/2008ApJ...679..346M} {679, 346}

\bibitem[\protect\citeauthoryear{{Napolitano}, {Capozziello}, {Romanowsky},
  {Capaccioli}  \& {Tortora}}{{Napolitano} et~al.}{2012}]{Napolitano2012}
{Napolitano} N.~R.,  {Capozziello} S.,  {Romanowsky} A.~J.,  {Capaccioli} M.,
  {Tortora} C.,  2012, \mn@doi [\apj] {10.1088/0004-637X/748/2/87}, \href
  {https://ui.adsabs.harvard.edu/abs/2012ApJ...748...87N} {748, 87}

\bibitem[\protect\citeauthoryear{{Navarro}, {Frenk}  \& {White}}{{Navarro}
  et~al.}{1996}]{Navarro1996}
{Navarro} J.~F.,  {Frenk} C.~S.,   {White} S. D.~M.,  1996, \mn@doi [\apj]
  {10.1086/177173}, \href
  {https://ui.adsabs.harvard.edu/abs/1996ApJ...462..563N} {462, 563}

\bibitem[\protect\citeauthoryear{{Nojiri} \& {Odintsov}}{{Nojiri} \&
  {Odintsov}}{2007}]{Nojiri2007}
{Nojiri} S.,  {Odintsov} S.~D.,  2007, \mn@doi [Physics Letters B]
  {10.1016/j.physletb.2007.10.027}, \href
  {https://ui.adsabs.harvard.edu/abs/2007PhLB..657..238N} {657, 238}

\bibitem[\protect\citeauthoryear{{Nojiri} \& {Odintsov}}{{Nojiri} \&
  {Odintsov}}{2011}]{Nojiri2011}
{Nojiri} S.,  {Odintsov} S.~D.,  2011, \mn@doi [\physrep]
  {10.1016/j.physrep.2011.04.001}, \href
  {https://ui.adsabs.harvard.edu/abs/2011PhR...505...59N} {505, 59}

\bibitem[\protect\citeauthoryear{{Nojiri}, {Odintsov}  \& {Oikonomou}}{{Nojiri}
  et~al.}{2017}]{Nojiri2017}
{Nojiri} S.,  {Odintsov} S.~D.,   {Oikonomou} V.~K.,  2017, \mn@doi [\physrep]
  {10.1016/j.physrep.2017.06.001}, \href
  {https://ui.adsabs.harvard.edu/abs/2017PhR...692....1N} {692, 1}

\bibitem[\protect\citeauthoryear{{Olszewski} \& {Aaronson}}{{Olszewski} \&
  {Aaronson}}{1985}]{Olszewski1985}
{Olszewski} E.~W.,  {Aaronson} M.,  1985, \mn@doi [\aj] {10.1086/113925}, \href
  {https://ui.adsabs.harvard.edu/abs/1985AJ.....90.2221O} {90, 2221}

\bibitem[\protect\citeauthoryear{{Peebles}}{{Peebles}}{2022}]{Peebles2022}
{Peebles} P. J.~E.,  2022, arXiv e-prints, \href
  {https://ui.adsabs.harvard.edu/abs/2022arXiv220805018P} {p. arXiv:2208.05018}

\bibitem[\protect\citeauthoryear{{Piatek}, {Pryor}, {Bristow}, {Olszewski},
  {Harris}, {Mateo}, {Minniti}  \& {Tinney}}{{Piatek}
  et~al.}{2007}]{Piatek2007}
{Piatek} S.,  {Pryor} C.,  {Bristow} P.,  {Olszewski} E.~W.,  {Harris} H.~C.,
  {Mateo} M.,  {Minniti} D.,   {Tinney} C.~G.,  2007, \mn@doi [\aj]
  {10.1086/510456}, \href
  {https://ui.adsabs.harvard.edu/abs/2007AJ....133..818P} {133, 818}

\bibitem[\protect\citeauthoryear{{Pietrzy{\'n}ski} et~al.,}{{Pietrzy{\'n}ski}
  et~al.}{2008}]{Pietrzynski2008}
{Pietrzy{\'n}ski} G.,  et~al., 2008, \mn@doi [\aj]
  {10.1088/0004-6256/135/6/1993}, \href
  {https://ui.adsabs.harvard.edu/abs/2008AJ....135.1993P} {135, 1993}

\bibitem[\protect\citeauthoryear{{Pietrzy{\'n}ski}, {G{\'o}rski}, {Gieren},
  {Ivanov}, {Bresolin}  \& {Kudritzki}}{{Pietrzy{\'n}ski}
  et~al.}{2009}]{Pietrzynski2009}
{Pietrzy{\'n}ski} G.,  {G{\'o}rski} M.,  {Gieren} W.,  {Ivanov} V.~D.,
  {Bresolin} F.,   {Kudritzki} R.-P.,  2009, \mn@doi [\aj]
  {10.1088/0004-6256/138/2/459}, \href
  {https://ui.adsabs.harvard.edu/abs/2009AJ....138..459P} {138, 459}

\bibitem[\protect\citeauthoryear{{Planck Collaboration} et~al.,}{{Planck
  Collaboration} et~al.}{2014}]{Planck2014}
{Planck Collaboration} et~al., 2014, \mn@doi [\aap]
  {10.1051/0004-6361/201321529}, \href
  {https://ui.adsabs.harvard.edu/abs/2014A&A...571A...1P} {571, A1}

\bibitem[\protect\citeauthoryear{{Planck Collaboration} et~al.,}{{Planck
  Collaboration} et~al.}{2020a}]{Planck2020-V}
{Planck Collaboration} et~al., 2020a, \mn@doi [\aap]
  {10.1051/0004-6361/201936386}, \href
  {https://ui.adsabs.harvard.edu/abs/2020A&A...641A...5P} {641, A5}

\bibitem[\protect\citeauthoryear{{Planck Collaboration} et~al.,}{{Planck
  Collaboration} et~al.}{2020b}]{Planck2020-VI}
{Planck Collaboration} et~al., 2020b, \mn@doi [\aap]
  {10.1051/0004-6361/201833910}, \href
  {https://ui.adsabs.harvard.edu/abs/2020A&A...641A...6P} {641, A6}

\bibitem[\protect\citeauthoryear{{Planck Collaboration} et~al.,}{{Planck
  Collaboration} et~al.}{2020c}]{Planck2020-VII}
{Planck Collaboration} et~al., 2020c, \mn@doi [\aap]
  {10.1051/0004-6361/201935201}, \href
  {https://ui.adsabs.harvard.edu/abs/2020A&A...641A...7P} {641, A7}

\bibitem[\protect\citeauthoryear{{Portinari}, {Sommer-Larsen}  \&
  {Tantalo}}{{Portinari} et~al.}{2003}]{Portinari2003}
{Portinari} L.,  {Sommer-Larsen} J.,   {Tantalo} R.,  2003, \mn@doi [\apss]
  {10.1023/A:1024060629820}, \href
  {https://ui.adsabs.harvard.edu/abs/2003Ap&SS.284..723P} {284, 723}

\bibitem[\protect\citeauthoryear{{Portinari}, {Sommer-Larsen}  \&
  {Tantalo}}{{Portinari} et~al.}{2004}]{Portinari2004}
{Portinari} L.,  {Sommer-Larsen} J.,   {Tantalo} R.,  2004, \mn@doi [\mnras]
  {10.1111/j.1365-2966.2004.07207.x}, \href
  {https://ui.adsabs.harvard.edu/abs/2004MNRAS.347..691P} {347, 691}

\bibitem[\protect\citeauthoryear{{Salucci} et~al.,}{{Salucci}
  et~al.}{2021}]{Salucci2021}
{Salucci} P.,  et~al., 2021, \mn@doi [Frontiers in Physics]
  {10.3389/fphy.2020.603190}, \href
  {https://ui.adsabs.harvard.edu/abs/2021FrP.....8..579S} {8, 579}

\bibitem[\protect\citeauthoryear{{Sohn} et~al.,}{{Sohn}
  et~al.}{2007}]{Sohn2007}
{Sohn} S.~T.,  et~al., 2007, \mn@doi [\apj] {10.1086/518302}, \href
  {https://ui.adsabs.harvard.edu/abs/2007ApJ...663..960S} {663, 960}

\bibitem[\protect\citeauthoryear{{Stabile} \& {Capozziello}}{{Stabile} \&
  {Capozziello}}{2013}]{2013PhRvD..87f4002S}
{Stabile} A.,  {Capozziello} S.,  2013, \mn@doi [\prd]
  {10.1103/PhysRevD.87.064002}, \href
  {https://ui.adsabs.harvard.edu/abs/2013PhRvD..87f4002S} {87, 064002}

\bibitem[\protect\citeauthoryear{{Starobinsky}}{{Starobinsky}}{1980}]{Starobinsky1980}
{Starobinsky} A.,  1980, Phys. Lett. B, 91, 99

\bibitem[\protect\citeauthoryear{{Starobinsky}}{{Starobinsky}}{2007}]{Starobinsky2007}
{Starobinsky} A.,  2007, JETP Lett., 86, 157

\bibitem[\protect\citeauthoryear{{The Theia Collaboration} et~al.,}{{The Theia
  Collaboration} et~al.}{2017}]{Theia2017}
{The Theia Collaboration} et~al., 2017, arXiv e-prints, \href
  {https://ui.adsabs.harvard.edu/abs/2017arXiv170701348T} {p. arXiv:1707.01348}

\bibitem[\protect\citeauthoryear{{Walker}, {Mateo}, {Olszewski}, {Gnedin},
  {Wang}, {Sen}  \& {Woodroofe}}{{Walker} et~al.}{2007}]{Walker2007}
{Walker} M.~G.,  {Mateo} M.,  {Olszewski} E.~W.,  {Gnedin} O.~Y.,  {Wang} X.,
  {Sen} B.,   {Woodroofe} M.,  2007, \mn@doi [\apjl] {10.1086/521998}, \href
  {https://ui.adsabs.harvard.edu/abs/2007ApJ...667L..53W} {667, L53}

\bibitem[\protect\citeauthoryear{{Walker}, {Mateo}  \& {Olszewski}}{{Walker}
  et~al.}{2009a}]{Walker2009a}
{Walker} M.~G.,  {Mateo} M.,   {Olszewski} E.~W.,  2009a, \mn@doi [\aj]
  {10.1088/0004-6256/137/2/3100}, \href
  {https://ui.adsabs.harvard.edu/abs/2009AJ....137.3100W} {137, 3100}

\bibitem[\protect\citeauthoryear{{Walker}, {Mateo}, {Olszewski}, {Sen}  \&
  {Woodroofe}}{{Walker} et~al.}{2009b}]{Walker2009b}
{Walker} M.~G.,  {Mateo} M.,  {Olszewski} E.~W.,  {Sen} B.,   {Woodroofe} M.,
  2009b, \mn@doi [\aj] {10.1088/0004-6256/137/2/3109}, \href
  {https://ui.adsabs.harvard.edu/abs/2009AJ....137.3109W} {137, 3109}

\bibitem[\protect\citeauthoryear{{Walker}, {Belokurov}, {Evans}, {Irwin},
  {Mateo}, {Olszewski}  \& {Gilmore}}{{Walker} et~al.}{2009c}]{Walker2009c}
{Walker} M.~G.,  {Belokurov} V.,  {Evans} N.~W.,  {Irwin} M.~J.,  {Mateo} M.,
  {Olszewski} E.~W.,   {Gilmore} G.,  2009c, \mn@doi [\apjl]
  {10.1088/0004-637X/694/2/L144}, \href
  {https://ui.adsabs.harvard.edu/abs/2009ApJ...694L.144W} {694, L144}

\bibitem[\protect\citeauthoryear{{Walker}, {Mateo}, {Olszewski},
  {Pe{\~n}arrubia}, {Evans}  \& {Gilmore}}{{Walker}
  et~al.}{2009d}]{Walker2009d}
{Walker} M.~G.,  {Mateo} M.,  {Olszewski} E.~W.,  {Pe{\~n}arrubia} J.,  {Evans}
  N.~W.,   {Gilmore} G.,  2009d, \mn@doi [\apj] {10.1088/0004-637X/704/2/1274},
  \href {https://ui.adsabs.harvard.edu/abs/2009ApJ...704.1274W} {704, 1274}

\bibitem[\protect\citeauthoryear{{Weinberg}}{{Weinberg}}{1987}]{Weinberg1987}
{Weinberg} S.,  1987, \mn@doi [\prl] {10.1103/PhysRevLett.59.2607}, \href
  {https://ui.adsabs.harvard.edu/abs/1987PhRvL..59.2607W} {59, 2607}

\bibitem[\protect\citeauthoryear{{Weinberg}}{{Weinberg}}{1989}]{Weinberg1989}
{Weinberg} S.,  1989, \mn@doi [Reviews of Modern Physics]
  {10.1103/RevModPhys.61.1}, \href
  {https://ui.adsabs.harvard.edu/abs/1989RvMP...61....1W} {61, 1}

\bibitem[\protect\citeauthoryear{{von Hoerner}}{{von
  Hoerner}}{1957}]{vonHoerner1957}
{von Hoerner} S.,  1957, \mn@doi [\apj] {10.1086/146321}, \href
  {https://ui.adsabs.harvard.edu/abs/1957ApJ...125..451V} {125, 451}

\makeatother
\end{thebibliography}

	


	
	\label{lastpage}
\end{document}